\documentclass[aip,pop,preprint,linenumber,superscriptaddress,amsmath,amsfonts]{revtex4-1}
\usepackage{color,graphicx} 
\usepackage{subfigure}%
\usepackage{lineno}

\usepackage[colorlinks,linkcolor=blue,citecolor=blue,bookmarks,pdfstartview=FitH]{hyperref}
\newcommand{\Poincare}{Poincar\'{e}\ }
\newcommand{\alfven}{Alfv\'en\ }

\begin{document}
	\title[Resonance frequency broadening]{Resonance frequency broadening of wave-particle interaction in tokamaks due to Alfv\'{e}nic eigenmode}
	\author{G. Meng}	
	\email{mengguo@pku.edu.cn}
	\affiliation{Fusion Simulation Center and State Key Laboratory of Nuclear Physics and Technology, School of Physics, Peking University, Beijing 100871, People's Republic of China}
	\affiliation{Princeton Plasma Physics Laboratory, Princeton University, Princeton, NJ 08543, United States of America}
	\author{N. N. Gorelenkov}
	\affiliation{Princeton Plasma Physics Laboratory, Princeton University, Princeton, NJ 08543, United States of America}
	\author{V. N. Duarte}
	\affiliation{Princeton Plasma Physics Laboratory, Princeton University, Princeton, NJ 08543, United States of America}
	\affiliation{Institute of Physics, University of S\~ao Paulo, S\~ao Paulo, SP, 05508-090, Brazil}
	\author{H. L. Berk}
	\affiliation{Institute for Fusion Studies, University of Texas, Austin, TX 78712, United States of America}
	\author{R. B. White}
	\affiliation{Princeton Plasma Physics Laboratory, Princeton University, Princeton, NJ 08543, United States of America}
	\author{A. Bhattacharjee}
	\affiliation{Princeton Plasma Physics Laboratory, Princeton University, Princeton, NJ 08543, United States of America}
	\author{X. G. Wang}
	\affiliation{Department of Physics, Harbin Institute of Technology, Harbin 150001, People's Republic of China}
	\affiliation{Fusion Simulation Center and State Key Laboratory of Nuclear Physics and Technology, School of Physics, Peking University, Beijing 100871, People's Republic of China}
	\date{\today}

	\begin{abstract}
	 We use a guiding center code ORBIT to study the broadening of resonances and the parametric dependency of the resonance frequency broadening width $\Delta\Omega$ on the nonlinear particle trapping frequency $\omega_b$ of wave-particle interaction with specific examples using realistic equilibrium DIII-D shot 159243 (Collins et al. 2016 Phys. Rev. Lett. \textbf{116} 095001). When the mode amplitude is small, the pendulum approximation for energetic particle dynamics near the resonance is found to be applicable and the ratio of the resonance frequency width to the deeply trapped bounce frequency $\Delta\Omega/\omega_b$ equals 4, as predicted by theory. This factor 4 is demonstrated for the first time in realistic instability conditions. It is found that as the mode amplitude increases, the coefficient $a=\Delta\Omega/\omega_b$ becomes increasingly smaller because of the breaking down of the nonlinear pendulum approximation for the wave-particle interaction. 
	\end{abstract}

	
	\maketitle

\section*{1. Introduction}
The confinement of energetic particles (EPs) is an important issue in present tokamaks and future fusion devices such as ITER limiting plasma operational regimes. EPs can drive shear Alfv\'enic\ gap modes and acoustic modes by releasing the free energy stored in the gradients of both real and velocity space. In turn, Alfv\'{e}nic eigenmodes (AEs) can modify the EP distribution through wave-particle interaction (WPI). These instabilities can cause significant losses of heat and EPs, which then downgrades confinement and even damages the tokamak walls \cite{nikolai14review}.\\
\indent The resonance condition which describes the interaction of eigenmodes and particles is fundamental to the mode growth rate. According to the quasilinear theory, diffusion only occurs for the particles that exactly satisfy the resonance condition via a delta function \cite{kaufman72QL}. However, the resonances are, in principle, broadened due to the finite growth rate and finite mode amplitude \cite{berk95LBQ}, and also effectively broadened due to stochastic processes affecting the resonant particles \cite{kattyLBQ}, such as collisions and microturbulence. Understanding the resonant broadened width of the WPI, affected by different dissipation mechanisms and nonlinear effects, can be crucial for the modeling of EP transport.\\
\indent When mode overlapping occurs, which implies that the Chirikov criterion \cite{chirikov79} is satisfied, then global diffusion of particles can occur. Particles move stochastically and therefore Conventional Quasilinear Theory (CQL) \cite{drummond&pines,vedenov61} provides a suitable description of the EP relaxation. We note, however, that the minimum stochastization of phase-space required by the CQL applicability can as well be provided by collisional \cite{kattyLBQ} and turbulent \cite{duarte17prediction} effects on resonant EPs. For the resonance overlapping situation, much of the fine phase-space structure of the distribution function is ``averaged out" due to particle decorrelation from a resonance. However, when individual resonances saturate at low levels without overlapping, they only perturb a localized portion of phase-space which does not lead to large-scale, global diffusion. In this case, particle redistribution can only occur within the resonance island because only the particles that are trapped by the wave undergo phase mixing. QL theory mimics the phase mixing by means of a diffusion coefficient that is not uniform in the phase-space, being non-zero only inside the resonance islands, where the distribution function is expected to be flattened. In addition, the particle diffusion due to a single mode is constrained to a one-dimensional manifold embedded in the full space.\\
\indent Recently, DIII-D experiments observed that the fast ion transport induced by Alfv\'{e}n eigenmode becomes stiff above a critical gradient. The results suggest that the threshold of AE-induced fast ion transport stiffness is above either the AE linear stability threshold or a microturbulent threshold \cite{collins16PRL}. To properly resolve stiff transport in the velocity space, reduced models accounting for transport thresholds that differ in phase-space are required. Its details need to be studied using realistic parameter of the fusion experiments. Recently, a promising computationally efficient code RBQ1D \cite{nikolaiRBQnf} was built based on the resonance broadening description by earlier publications \cite{berk96nonlinear,kattyLBQ}. It employs the Resonance Broadening Quasilinear (RBQ) model \cite{berk95LBQ} to study the EP transport including phase-space dependence of the diffusion coefficient and the mode amplitude self-consistent evolution with time. This reduced model and can handle both isolated or overlapped modes \cite{nikolaiRBQnf}. In this work, we use the guiding center test particle code ORBIT \cite{white84ORBIT,white2013TCP} to test assumptions of the RBQ model regarding the width of the platform for WPI. Our goal is to find the parametric dependence of the resonance broadening and discuss the validity of this model in a realistic DIII-D case such as the one described in recent work \cite{collins16PRL}.\\ 
\indent This paper is organized as follows. In Section 2, we introduce the RBQ model for resonance line broadening. In Section 3, we propose a new numerical approach to measure the resonance frequency broadening due to a single mode. In Section 4, we propose a numerical method to calculate the trapping frequency of WPI in a particle following code exploiting kinetic \Poincare plots. In Section 5, we investigate the resonance broadening in four realistic DIII-D cases and discuss the break-down of the nonlinear pendulum approximation.
\section*{2. Resonance Line Broadening Formalism}
In tokamaks, an unperturbed particle trajectory is periodic in three canonical angles and can be described by three constants of motion (COM): particle energy $\mathcal{E}$, toroidal canonical momentum $P_\zeta$ and magnetic moment $\mu$. In the guiding center code ORBIT, the time is normalized with $\omega_{0}^{-1}$, where $\omega_{0}=Z_{EP}eB_0/m_{EP}c$ is the on-axis gyro-frequency of EP, $B_0$ the on-axis magnetic field strength, $Z_{EP}$ the EP atomic number, $e$ the elementary charge and $m_{EP}$ the EP mass, $c$ the light speed; the length is normalized with the major radius $R_0$. The unperturbed Hamiltonian is given by
\begin{equation}
H_0=\frac{(\rho_{\parallel}B)^2}{2}+\mu B+\Phi,
\end{equation}
where $\rho_{\parallel}=v_\parallel/B$ is the ``parallel gyro radius", and $\Phi$ is the electric potential. The guiding center equations used in ORBIT advance particle variables $(\psi_p,\theta,\zeta,\rho_{\parallel})$, leaving $\mu$ as a constant of the motion. Equilibrium field quantities are given by $B=g\nabla\zeta+I\nabla \theta+\delta\nabla\psi_p$ with $\psi_p$ the poloidal flux over $2\pi$, $\theta$ the poloidal angle, $\zeta$ a toroidal angle coordinate, and $g(\psi_p)$, $I(\psi_p)$ and $\delta(\psi_p,\theta,\zeta)$ are equilibrium functions. The toroidal and poloidal canonical momenta are 
\begin{equation}\label{eq:Pphi}
P_\zeta=g\rho_{\parallel}-\psi_p,\qquad P_\theta=\rho_{\parallel}+\psi_t,
\end{equation}
where $\psi_t$ is the toroidal flux over $2\pi$, and $q(\psi_p)=d\psi_t/d\psi_p$ is the safety factor \cite{white84ORBIT,white2013TCP}. In the rest of this paper, $\psi_p$ refers to the normalized poloidal flux $\psi_p/\psi_w$, where $\psi_w$ is value of poloidal flux at the plasma edge. The EP unperturbed distribution can be written as $f_0(\mathcal{E},P_\zeta,\mu)$, which can be estimated from a theoretical description or found more precisely using a numerical beam deposition code, such as NUBEAM routine of TRANSP \cite{transp04}.
Oscillations are prescribed through the following ansatz for the fluid displacement:
\begin{equation}
\vec{\xi}=\sum_{n,m}A_n\xi_{n,m}(\psi_p)e^{-i(n\zeta-m\theta-\omega_n t)},
\end{equation} 
where $n$ refers to the toroidal mode number for a single mode with angular frequency $\omega_n$, with several poloidal harmonics $m$ and the amplitude $A_n$ is the magnitude of the ideal displacement caused by this mode. The Hamiltonian is a function of $n\zeta-\omega_n t$ if there is only a single mode. Then we have
\begin{equation}
\dot{P_\zeta}=-\frac{\partial H}{\partial\zeta}=n\dot{H}/\omega_n,
\end{equation} 
and thus $\mathcal{E}'=\mathcal{E}-\omega_n P_\zeta/n=const$ for a single mode (the value of the Hamiltonian is the energy $\mathcal{E}$). The motion of particles in the $(\mathcal{E},P_\zeta)$ plane is therefore restricted to one dimension.\\
\indent When the mode frequency is much lower than the cyclotron frequency $\omega_n\ll\omega_c$, the wave-particle linear resonance condition is given by
\begin{equation}\label{eq:res}
\Omega(\mathcal{E},P_\zeta,\mu)=n\langle\omega_\zeta\rangle-l\langle\omega_\theta\rangle-\omega_n=0,
\end{equation} 
where $l$ is an integer, $\omega_\zeta$ and $\omega_\theta$ are the toroidal and poloidal transit frequencies and $\langle\ldots\rangle$ denotes averaging over one poloidal transit time \cite{berk96nonlinear}. Note that in this expression the integer $l$ is not the poloidal mode number $m$. Usually the integer $l$ is close to the poloidal mode number $m$ existing in the mode, but the exact resonance location can be only found by integrating over the particle poloidal trajectory \cite{white2013TCP}.\\  
\indent Although the linear resonance condition implies that the Eq. \ref{eq:res} is only satisfied at singular points, there are effects that can naturally broaden the resonance, e.g., via nonlinear \cite{Dupree66,kattyLBQ} and collisional \cite{BB90II} mechanisms.  The RBQ model \cite{berk95LBQ,berk96nonlinear} was conceived as a modification of the usual quasilinear relaxation framework that copes with the expected single-mode saturation levels, as predicted by analytical theory. The RBQ employs the same structure of the quasilinear theory while considering that the resonances are broadened around a central point determined by the linear resonance condition in Eq. \ref{eq:res} \cite{nikolaiRBQnf}. The broadened resonance is the platform that allows momentum and energy exchange between particles and waves. It is assumed that the width of frequency broadening around resonance $\Omega=0$ is \cite{berk95LBQ}
\begin{equation}
\Delta\Omega=a\omega_b+b\gamma+c\nu_{eff},
\end{equation} 
where $a$, $b$ and $c$ are coefficients, $\omega_b$ is the deeply-trapped particle bounce frequency of WPI, $\nu_{eff}$ is the effective scattering rate, and $\gamma$ is the net growth rate. The coefficients $a$ and $c$ in two extreme limits (far from and close to marginal stability) can be calculated analytically, respectively \cite{BB90I,BB90II,BB90III,kattyLBQ}. In addition, the saturation level of toroidicity-induced \alfven eigenmode (TAE) in TFTR of numerical simulation and the prediction from single wave saturation are in good agreement \cite{nikolai99saturation}. In this work, we study the resonance broadening due to a single mode to validate the coefficient $a$. A consistent study of mode saturation level to calculate coefficients $b$ and $c$ will be the scope of further work.
\section*{3. Numerical Approach}
In this work, we propose a numerical approach to measure the resonance broadening width and the trapping frequency. This method could in principle be used for detailed study of the distribution function modification due to WPI.
\subsection*{3.1. Equilibrium and Mode}
\begin{figure*}\centering	
	\includegraphics[width=3in, angle=0]{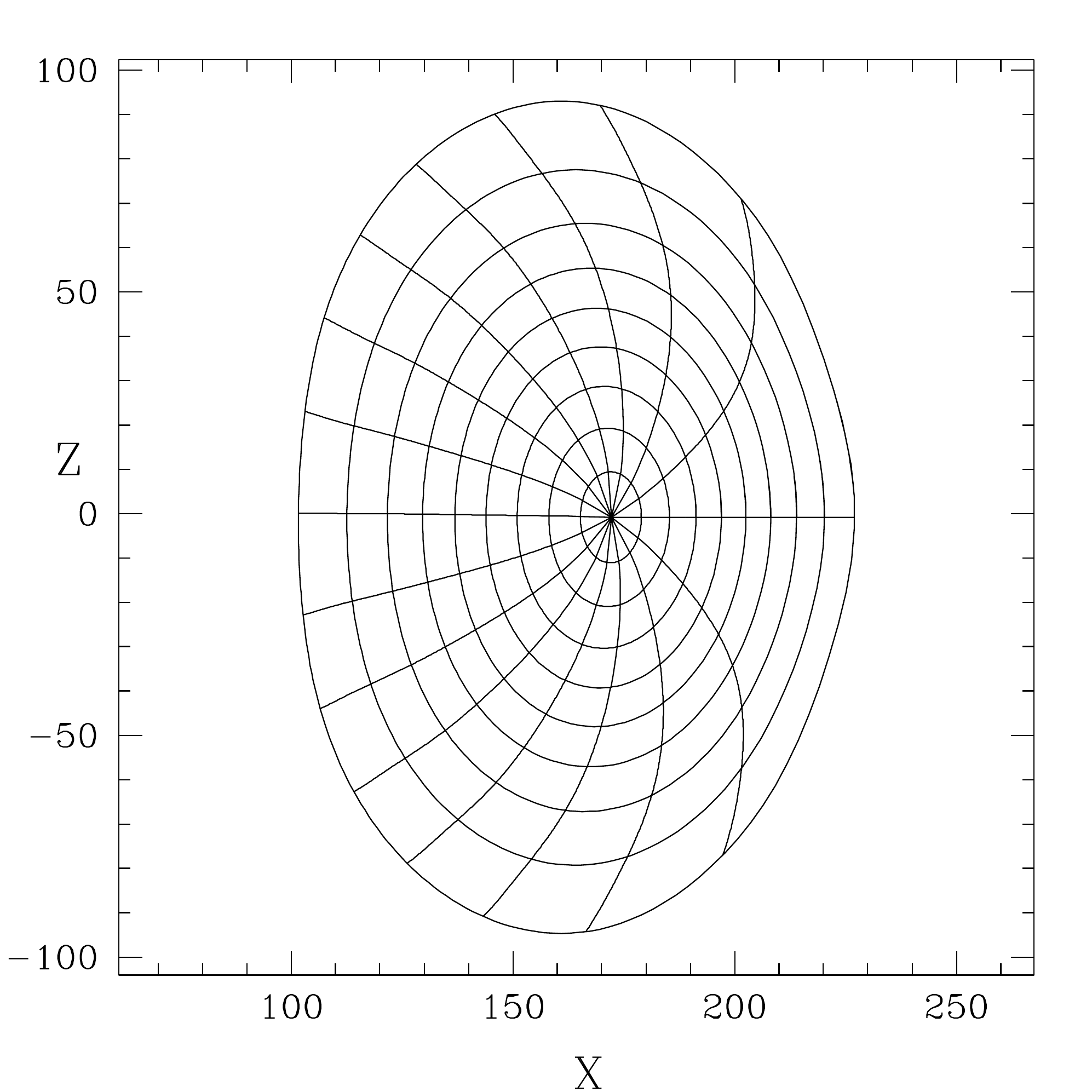}
	\caption{Plasma equilibrium shown in a poloidal cross section with X and Z given in centimeters.}
	\label{fig:equilibrium}
\end{figure*}

We use the parameters of the DIII-D discharge 159243 at 790 ms. This case is described in Collins' paper on the EP critical gradient model \cite{collins16PRL}. Figure \ref{fig:equilibrium} shows the equilibrium for this case. Other parameters are $B_0=19.77\;kG$, major radius $R_0=172\;cm$, minor radius $a=63\;cm$, $q_{min}=2.95$ at $\psi_p=0.24$.
\begin{figure*}\centering	
	\includegraphics[width=3.2in, angle=0]{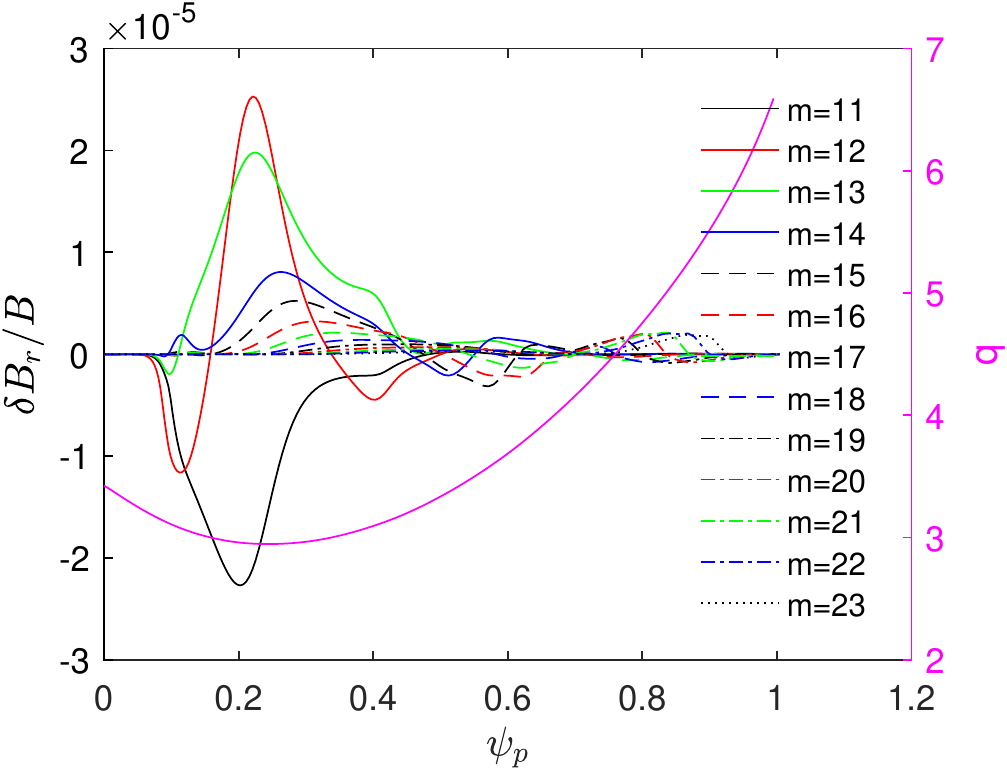}
	\caption{The left axis shows the mode structure of the $n=4$ RSAE with mode frequency 82.03 kHz. The normalized mode amplitude is the magnetic perturbation of the dominant harmonic $m=12$ and $A=\delta B_r/B=2.5\times10^{-5}$. The right axis shows the safety factor $q(\psi_{p})$, which has a minimum around the mode structure peak.}
	\label{fig:mode}                
\end{figure*}
To carry out simulations of realistic discharges, we use a mode structure and frequency determined by the ideal MHD code NOVA \cite{nova86}. The chosen Reversed Shear Alfv\'enic\ Eigenmode (RSAE) near $q_{min}$ is critical to understand the EP transport leading to hollow EP pressure profiles, and yet the case we consider is nevertheless general to make conclusions for future RBQ development. The mode structure of this RSAE is shown in Fig. \ref{fig:mode} with a single toroidal mode number $n=4$ and frequency 82.03 kHz, with poloidal harmonics $m$ ranging from 11 to 23. The mode peaks at $\psi_p=0.22$ near the $q_{min}$ and the dominant poloidal harmonic is $m=12$. We use the largest value of the dominant poloidal component m=12 as the mode amplitude $A=\delta B_r/B$. 
\subsection*{3.2. NBI particle distribution}
\begin{figure*}\centering
	\subfigure{\label{fig:fEpitch}\includegraphics[width=3in, angle=0]{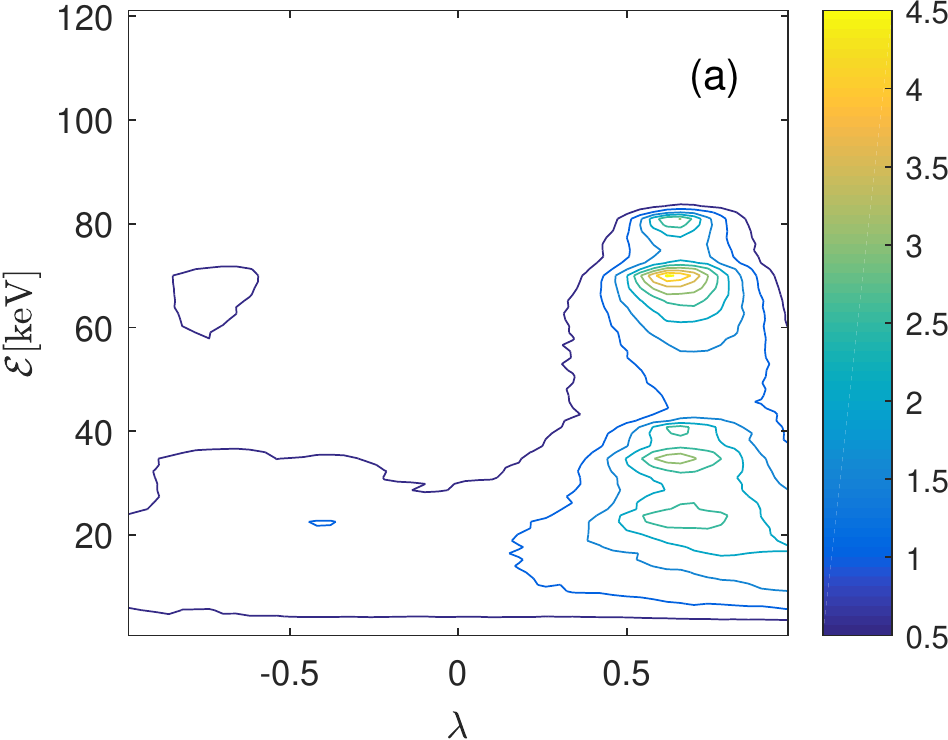}}
	\subfigure{\label{fig:fmuB}\includegraphics[width=2.71in, angle=0]{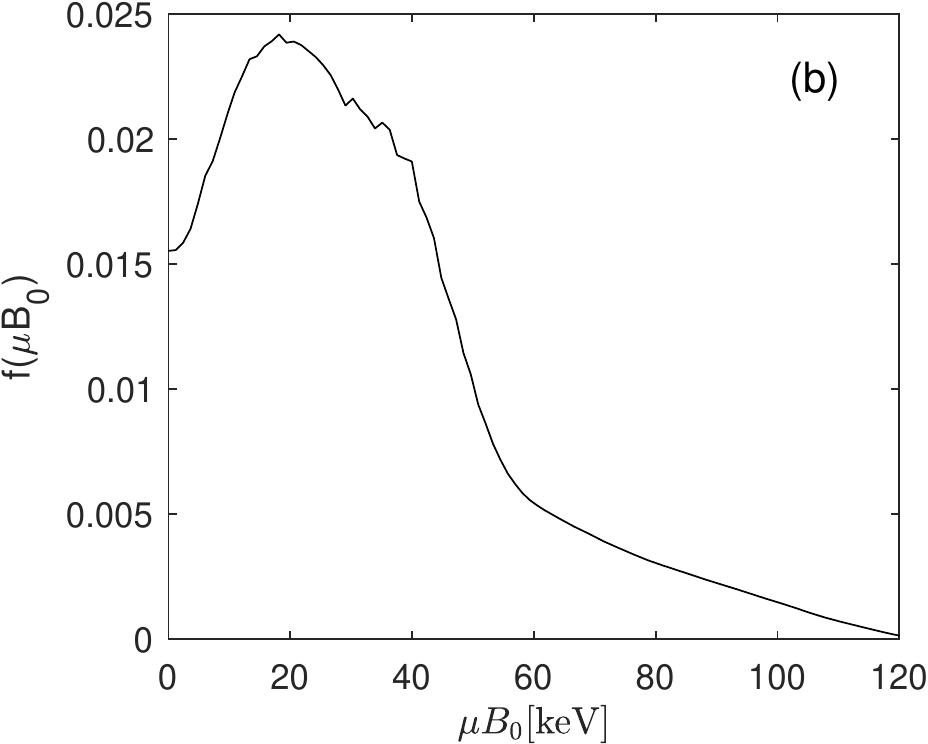}}
	\caption{EP distribution for DIII-D shot 159243 at $790\;ms$ obtained from NUBEAM simulation. (a): EP distribution of energy and pitch. (b): EP distribution in term of the magnetic moment.}
	\label{fig:NBI}                
\end{figure*}

EP distributions are obtained from the beam deposition code NUBEAM in TRANSP \cite{transp04}. The distribution function $f(\mathcal{E},\lambda)$ is integrated over the minor radius as shown in Fig. \ref{fig:fEpitch}, where $\lambda=v_\parallel/v$ is the pitch. This discharge had tangential co-injection and the injected energy is $\mathcal{E}_{inj}=70\;keV$. Figure \ref{fig:fmuB} shows the distribution in terms of the magnetic moment $f(\mu B_0)$, which is integrated over the other coordinates. Without collisions, $\mu$ is a constant for the guiding center equations used in ORBIT. We use $\mu B_0$ to convert the value of the magnetic moment to units of energy. The distribution $f(\mu B_0)$ peaks at $\mu B_0=20\;keV$, so we choose co-moving particles in the $(\mathcal{E},P_\zeta,\mu B_0=20\;keV)$ plane for detailed studies.
\subsection*{3.3. Wave-Particle Resonance}
\begin{figure*}\centering
	\subfigure
	{\label{fig:OmegaPEplane} 
		\includegraphics[width=5in, angle=0]{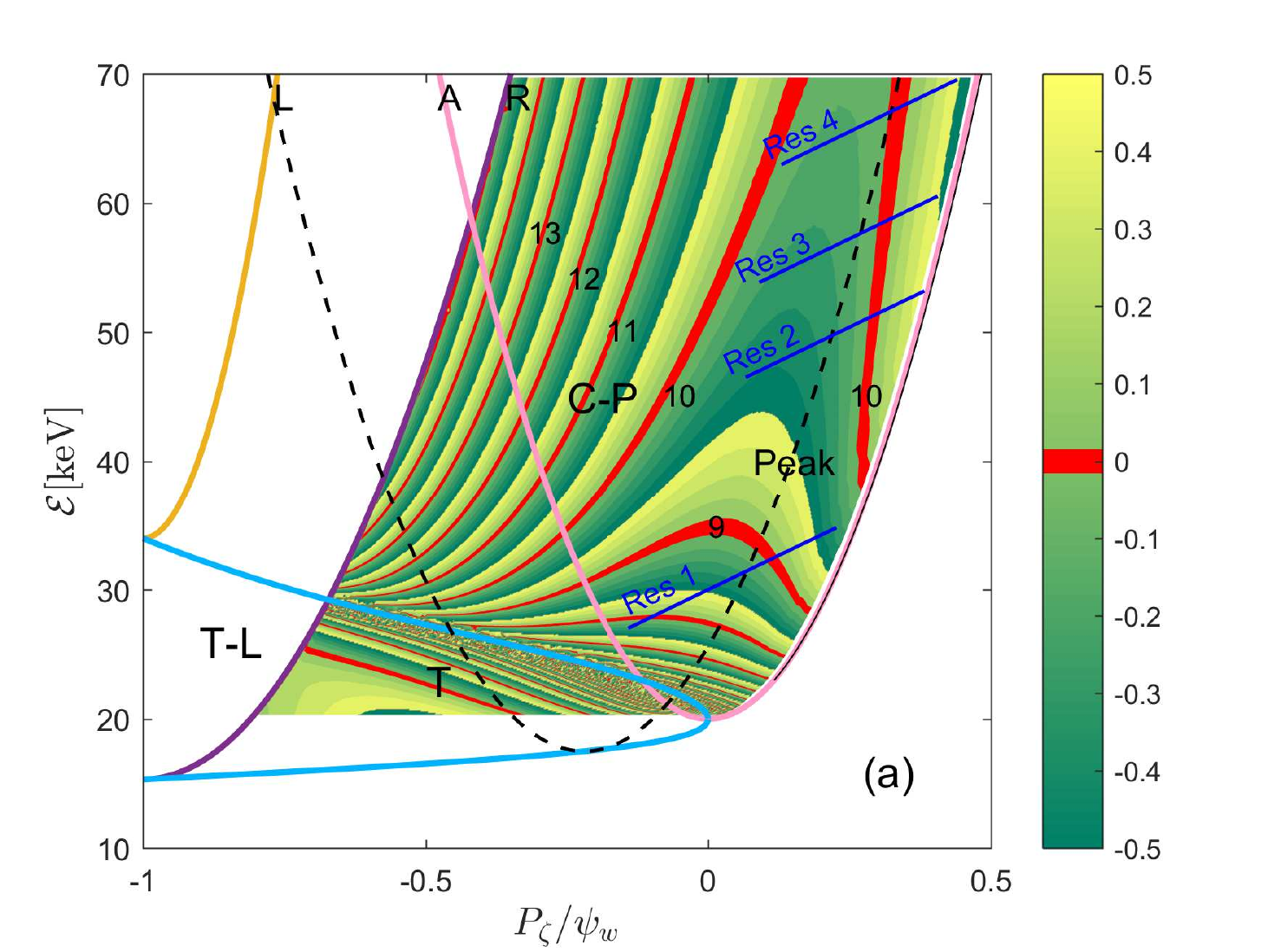}}
	\subfigure
	{\label{fig:wormamp6}
		\includegraphics[width=3.7in, angle=0]{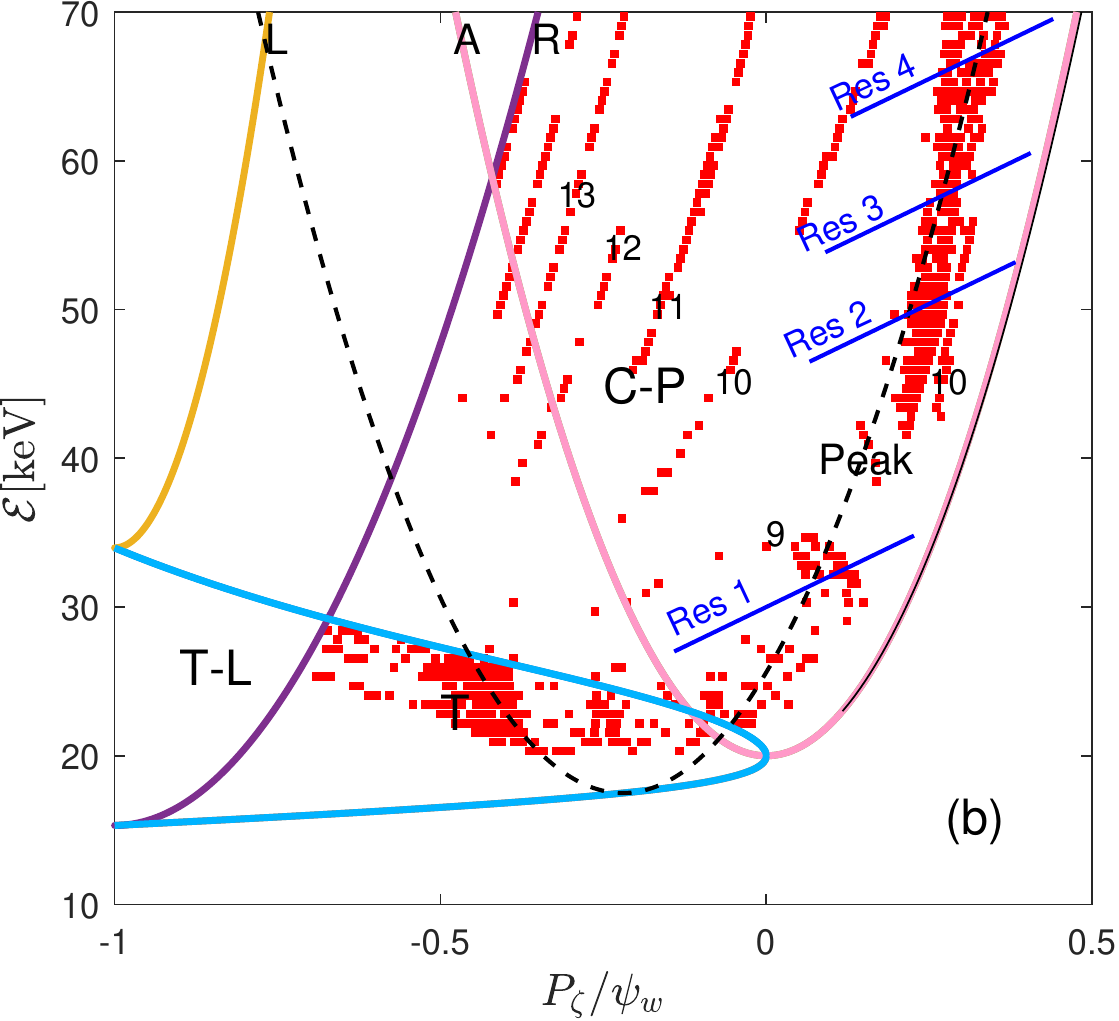}}
	\caption{Resonance slice at the $\mu B_0=20\;keV$ plane. (a): Colorbar value equals to $\Omega/\omega_\theta$ and indicates the strength of resonance. Red lines show resonances condition being satisfied, $\Omega=0$, with $l=9,10,11$ and so on. (b): Broken KAM surface domains are indicated by the red dots, which are detected by phase vector rotation at mode amplitude $A=1.5\times10^{-4}$.\\
		Four parallel {\color{blue}blue} lines from bottom to top are initial particle deposit positions Res 1, 2, 3, 4 with $\mathcal{E}'=30\;keV,45.09\;keV,51.93\;keV,60.22\;keV$, respectively. Also, trapped-loss (T-L), trapped (T), and co-passing (C-P) domains are shown. Particle orbits intersecting with the magnetic axis (A), the right wall (R) and the left wall (L) are shown with pink, purple and orange lines, respectively. The dashed black line shows particle orbit passing through the $(\psi_p=0.22,\theta=0)$ point, where the mode structure peaks.}
	\label{fig:resPEplane}                
\end{figure*}

Linear wave-particle resonance $\Omega=0$ can be calculated numerically by following particle trajectories in equilibrium, i.e., in the \textit{absence} of a mode. The toroidal and poloidal transit frequencies $\omega_\zeta$ and $\omega_\theta$ are calculated by following 50 000 particles and averaging the frequencies over about 10 poloidal transit times, in order to achieve reasonable accuracy. Then we use cubic interpolation to get a smooth resonance frequency plot of 200 000 points uniformly distributed on the $(\mathcal{E},P_\zeta)$ plane. The interpolation is used to calculate $\langle\omega_\zeta\rangle$ and $\langle\omega_\theta\rangle$ and then $\Omega$ can be readily computed. \\
\indent For calculating the resonance frequency, we first calculate $(n\langle\omega_\zeta\rangle-\omega_n)/\langle\omega_\theta\rangle$ to find the integer $l$ on this $(\mathcal{E},P_\zeta)$ plane, where $l$ is the closest integer to $(n\langle\omega_\zeta\rangle-\omega_n)/\langle\omega_\theta\rangle$. As shown in Fig. \ref{fig:OmegaPEplane}, the colorbar value equals to $\Omega/\omega_\theta$ and indicates the location of resonance. When $\Omega/\omega_\theta=0$, linear resonance condition is satisfied and $\left| \Omega/\omega_\theta \right| =0.5$ is far from resonance. Red lines are where the resonances conditions $\Omega=0$ are satisfied with $l=9,10,11$ and so on. For co-passing particles, $\omega_\zeta\approx q\omega_\theta$, the resonance lines have a turning point somewhere in $(\mathcal{E},P_\zeta)$ plane for reversed $q$ profile. This property can be clearly seen from the $l=9$ resonance. For the $l=10$ resonance, its turning point is at higher energy and outside the region $20\;keV<\mathcal{E}<70\;keV$ we studied. Even though there are many resonances near the trapped-passing separatrix, they do not contribute much to the growth rate.\\ 
\indent The resonance lines show the linear resonance location in phase-space but they do not provide information on the resonance island width or on the strength of the local phase-space contribution to the growth rate, which is quite sensitive to the local distribution function. The phase vector rotation technique \cite{white12PhaseVector,white11ppcf} embedded in ORBIT code can be used to detect the broken Kolmogorov-Arnold-Moser (KAM) \cite{K1954,A1963,M1962} surfaces efficiently and accurately in the \textit{presence} of a single mode. The red points indicate the location of the broken KAM surfaces for a mode amplitude $A=1.5\times10^{-4}$. In Fig. \ref{fig:wormamp6} we can see that there is a dominant resonance near the right boundary, which corresponds to the magnetic axis. The dominant resonance is located near the largest mode amplitude, where the mode structure peaks.
\subsection*{3.4. Kinetic \Poincare plot}
\begin{figure*}\centering
	\subfigure
	{\label{fig:poink1} 
		\includegraphics[width=3in, angle=0]{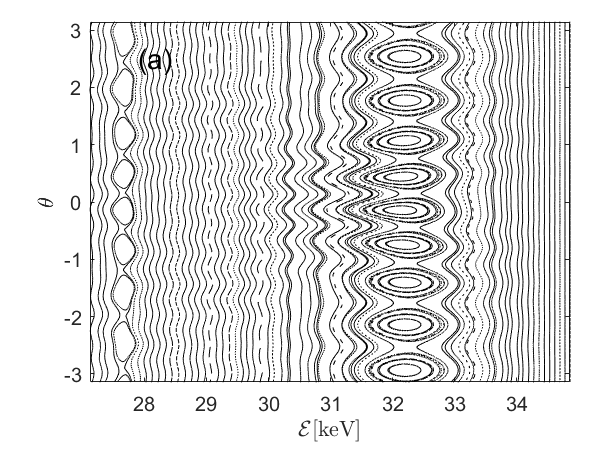}}
	\subfigure
	{\label{fig:poink2}
		\includegraphics[width=3in, angle=0]{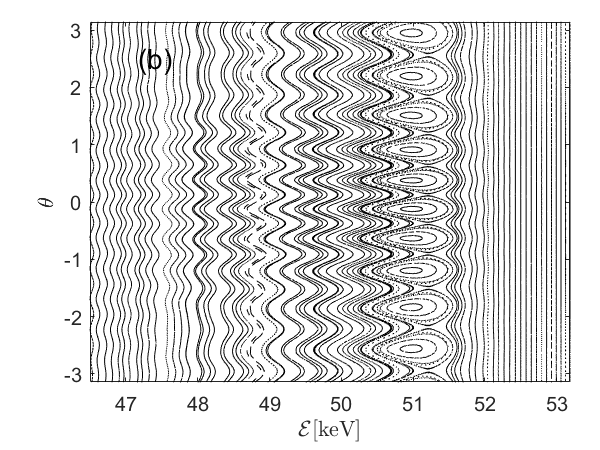}}
	\subfigure
	{\label{fig:poink3} 
		\includegraphics[width=3in, angle=0]{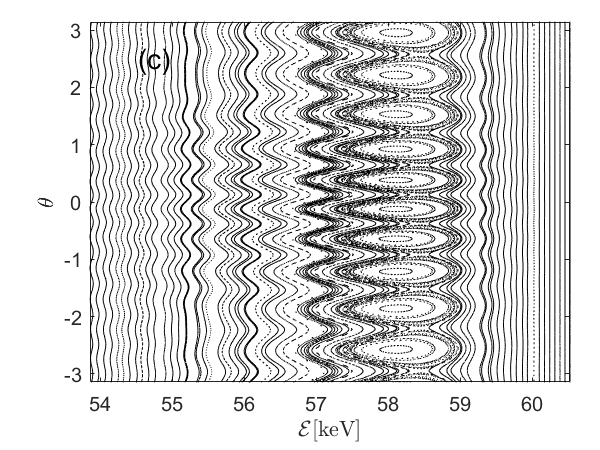}}
	\subfigure
	{\label{fig:poink4} 
		\includegraphics[width=3in, angle=0]{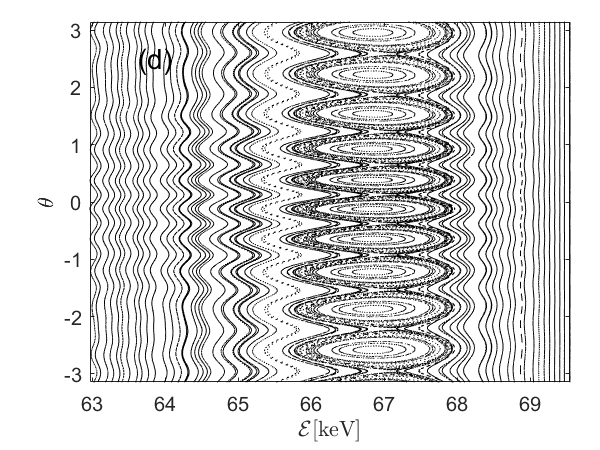}}
	\caption{Kinetic \Poincare plots at these four resonance locations (blue lines in Fig. \ref{fig:resPEplane}), with same mode amplitude $A=1.5\times10^{-4}$. (a): Res 1, $\mathcal{E}'=30\;keV$, $l=9$. (b): Res 2, $\mathcal{E}'=45.09\;keV$, $l=10$. (c): Res 3, $\mathcal{E}'=51.93\;keV$, $l=10$. (d): Res 4, $\mathcal{E}'=60.22\;keV$, $l=10$.}
	\label{fig:poinkall}                
\end{figure*}

For more detailed study on the wave-particle resonance, a kinetic \Poincare plot can be employed to provide the high-resolution figures of the resonance island structure \cite{white2013TCP}. Contrary to the resonance condition plot Fig. \ref{fig:OmegaPEplane}, kinetic \Poincare plots are produced by following particle orbits in the \textit{presence} of a mode with a single toroidal mode number and frequency, and recording points whenever $n\zeta-\omega t=2k\pi$ with $k$ integer. 
In addition, when an island overlaps with other islands, the kinetic \Poincare plot become chaotic, which implies that the Chirikov criterion \cite{chirikov79} is satisfied for this situation. The kinetic \Poincare plot shows the resonance only along the line $\mathcal{E}^{\prime}=constant$, with fixed magnetic moment $\mu$. For studying the dominant resonance we are interested in, we choose four cases with $\mathcal{E}^{\prime}=30\;keV,45.09\;keV,51.93\;keV,60.22\;keV$ intersecting with the dominant resonance lines, respectively. Their locations are the four parallel {\color{blue}blue} lines as shown in Fig. \ref{fig:resPEplane} from bottom to top. We will refer to them as Res 1, Res 2, Res 3, Res 4. The choice of these four lines is sufficiently general. We cover the dominant contributions to the mode growth rate which show the variation of the nonlinear pendulum description to resonance dynamics.\\
\indent Figure \ref{fig:poinkall} shows the kinetic \Poincare plots for these four cases with mode amplitude $A=1.5\times10^{-4}$. The integer $l$ in resonance frequency $\Omega$ represents the periodicity of the resonance island chain, which means $l$ islands in the $\theta$ direction as seen in the kinetic \Poincare plots. For Res 1, $l=9$ as shown in Fig. \ref{fig:poink1}. For Res 2, 3 and 4, $l=10$ as shown in Figs. \ref{fig:poink2}, \ref{fig:poink3} and \ref{fig:poink4}, respectively.
\section*{4. Wave-Particle Trapping Frequency}
The trapping frequency $\omega_{bt}$ is the frequency of the bounce motion of a resonant particle trapped by the wave. It is a function of distance from the island O-point, dropping to zero at the separatrix. The deeply trapped bounce frequency $\omega_b$ is $\omega_{bt}$ at the O-point, which is proportional to the width of the island and to the square root of mode amplitude, $\omega_b\propto A^{1/2}$. The wave-particle trapping frequency for small amplitude oscillations, i.e., of a particle near O-point is given by \cite{berk97report} 
\begin{equation}\label{eq:wb}
\omega_b^2=\left\langle Z_{EP} e\bm{E}\cdot \bm{v} \frac{n}{\omega} \frac{\partial \Omega}{\partial P_\zeta}\bigg|_{\mathcal{E}'}\right\rangle, 
\end{equation}
where $\bm{E}$ is the electric field, $\bm{v}$ is the velocity, and the time average is taken for an exactly resonant particle. \\ 
\indent We observe that, the resonance islands are wider at the dominant resonance, where particle orbits intersect with the location of the mode peak as inferred from Fig. \ref{fig:wormamp6}. Outside of this region, resonance islands are relatively narrow, which means small $\omega_{b}$. From Res 1 to Res 4, the resonance islands become increasingly stretched along the energy coordinate at the same mode amplitude, as inferred from Fig. \ref{fig:poinkall}. It implies that the deeply trapped bounce frequency $\omega_{b}$ is increasing, which can be seen from Fig. \ref{fig:allReswb_A}. The results show the $\omega_{b}$ is larger at the dominant resonance and high-energy regime.\\
The trapping frequency $\omega_{bt}$ is the angular frequency of a particle rotating $2\pi$ in the island as shown in the kinetic \Poincare plot. In this paper, we propose a numerical approach to calculate the trapping frequency and developed a MATLAB package for ORBIT data analysis. The trapping frequency of a particle is calculated numerically by following its orbit on the kinetic \Poincare plot. The details can be found in Appendix A.
\begin{figure*}\centering
	\subfigure
	{\label{fig:poinken} 
		\includegraphics[width=1.95 in, angle=0]{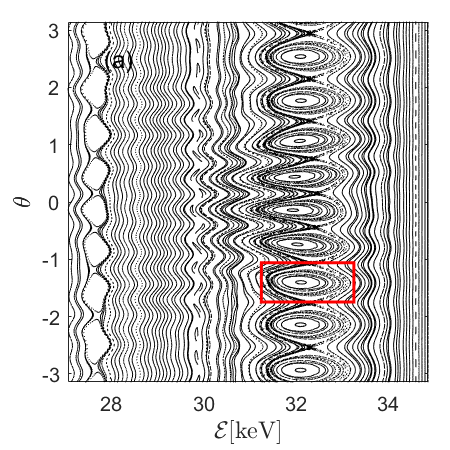}}
	\subfigure
	{\label{fig:island}
		\includegraphics[width=1.95 in, angle=0]{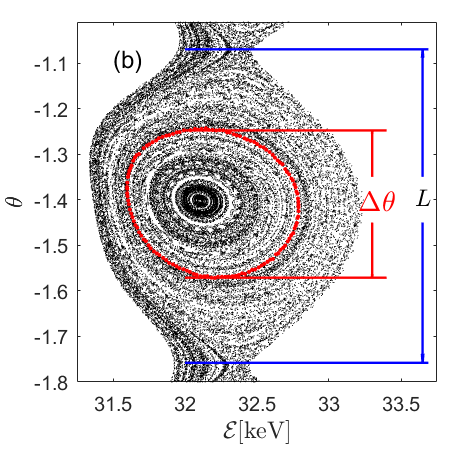}}             
	\subfigure
	{\label{fig:wb}
		\includegraphics[width=1.95 in, angle=0]{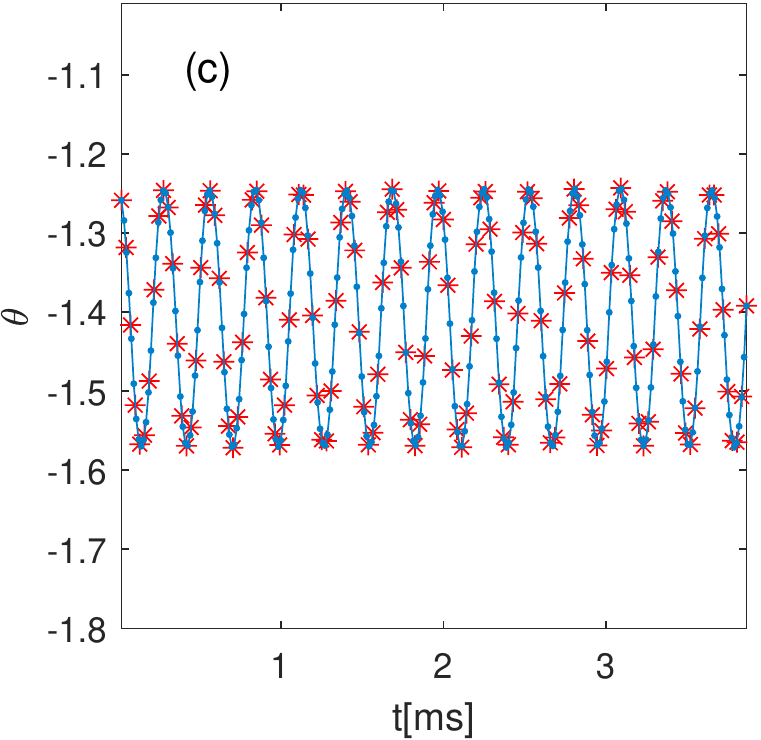}}
	\caption{Example of trapping frequency calculation with a mode amplitude $A=3.25\times10^{-4}$ at Res 1. (a): Kinetic \Poincare plot. The resonance island in the red rectangle is chosen to calculate the trapping frequency and ``blown up" in (b). (b): Definition of $\Delta\theta$ and island length $L=2\pi/l$. The orbit of one particle is shown in red and its $\theta$ variation with time is shown in (c). The trapping frequency of this particle is $\omega_{bt}=22.89\;rad/ms$.}\label{fig:Egwb}                
\end{figure*}
\subsection*{4.1. Pendulum Approximation}
Since the resonance islands are stretched increasingly with increasing mode amplitude, highly asymmetric in $\mathcal{E}$ direction, and particle orbits are stochastic near separatrix, it is hard to identify the island width accurately (see Fig. \ref{fig:deform}). However, the resonance islands are nearly symmetric and do not stretch with mode amplitude in $\theta$ direction, so we choose $\Delta\theta$ as the variable to represent the distance from O-point. So we define the resonance island non-dimensional length in $\theta$ direction, which is an action angle. Then, the resonance island length is $L=2\pi/l$ as shown in Figs. \ref{fig:poinken} and \ref{fig:island}. The red rectangle is the resonance island chosen to calculate the trapping frequency. \\
For the case of a sinusoidal wave, the collisionless motion of the resonant particle around the O-point satisfies the pendulum equation \cite{berk97report}
\begin{equation}\label{eq:pendulum}
\frac{d^2\alpha}{dt^2}+\omega_b^2sin(\alpha-\omega_{n}t-\alpha_0)=0,
\end{equation}
where $\alpha$ is the action angle and $\alpha_0$ is a constant phase.
\begin{equation}\label{eq:actionangle}
\alpha=(n,-l,0)\cdot(\zeta,\theta,\theta_c)=n\zeta-l\theta,
\end{equation} 
where $n$, $-l$, and $0$ are the quantum numbers associated with the periodicity of the canonical angles $\zeta$, $\theta$ and $\theta_c$, the gyroangle, so the resonance frequency is $\Omega= \dot{\alpha} -\omega_n$. Similarly to a pendulum, the energy-like variable of a trapped particle in the wave frame is
\begin{equation}
E_{bt}=\dot{\alpha}^2/2+\omega_b^2[1-cos(\alpha-\omega_{n}t-\alpha_0)],
\end{equation} 
where the first term and the second term represent the kinetic and the potential energies of the pendulum, respectively. The trapped particle energy ranges from 0 to $E_{bt}=2\omega_b^2$, with $E_{bt}=0$ at the O-point and $E_{bt}=2\omega_b^2$ at the separatrix. The value of $\alpha$ changes with time from $-\pi$ to $\pi$ when the particle orbit is at the separatrix, so we have
\begin{equation}
\dot{\alpha}^2/2=\omega_b^2[1+cos(\alpha-\omega_{n}t-\alpha_0)]\leq2\omega_b^2.
\end{equation}
Therefore,
\begin{equation}
\Delta\Omega=\Omega_{max}-\Omega_{min}=4\omega_b,
\end{equation}
for trapped particles, which is the predicted resonance broadening width. Since, to the best of our knowledge, the factor 4 has not yet been checked in a realistic case, we perform a detailed study of the pendulum approximation in this paper.\\ 
The condition for a kinetic \Poincare plot is $n\zeta-\omega_nt=2k\pi$. Substituting this condition and Eq. \ref{eq:actionangle} into Eq. \ref{eq:pendulum}, then we have
\begin{equation}
-\frac{d^2(l\theta)}{dt^2}+\omega_b^2sin(l\theta+\alpha_0)=0.
\end{equation}  
So we can derive the nonlinear trapping frequency when the pendulum approximation for the WPI is valid
\begin{equation}\label{eq:wbt}
\omega_{bt}=\omega_b\frac{\pi}{2}\frac{1}{K(k)},
\end{equation}
where $K(k)=\int_0^\frac{\pi}{2} \frac{1}{\sqrt{1-k^2\sin^2 u}}\,du\,$ is the first kind complete elliptic integral, $k=sin(\frac{\pi}{2}\overline{\Delta\theta})$, $\overline{\Delta\theta}=\Delta\theta/L$ is the normalized orbit length of a trapped particle and the angular range $\Delta\theta$ is indicated in Fig. \ref{fig:island}. The deeply trapped bounce frequency is $\omega_{b}=\omega_{bt}(\Delta\theta=0)$. Since the trapping frequency changes slowly in the island center, numerically we set $\omega_b$ equal to the trapping frequency of the particle with the minimal $\Delta\theta$. Figure \ref{fig:island} presents the selected resonance island as shown in Fig. \ref{fig:poinken}, which is ``blown up'' by loading more particles in the resonance. The orbit of one particle is highlighted in red. We also plot its action angle $\theta$ changing with time in Fig. \ref{fig:wb}. The red star points are numerical data and the blue dots are interpolated points to improve the accuracy of Fast Fourier Transform (FFT). Then we use FFT to analyze the signal $\theta(t)$ to calculate the trapping frequency. The details can be found in Appendix A.
\subsection*{4.2. Numerical Method to Measure the Resonance Broadening}
\begin{figure}\centering
	\includegraphics[width=0.6\textwidth]{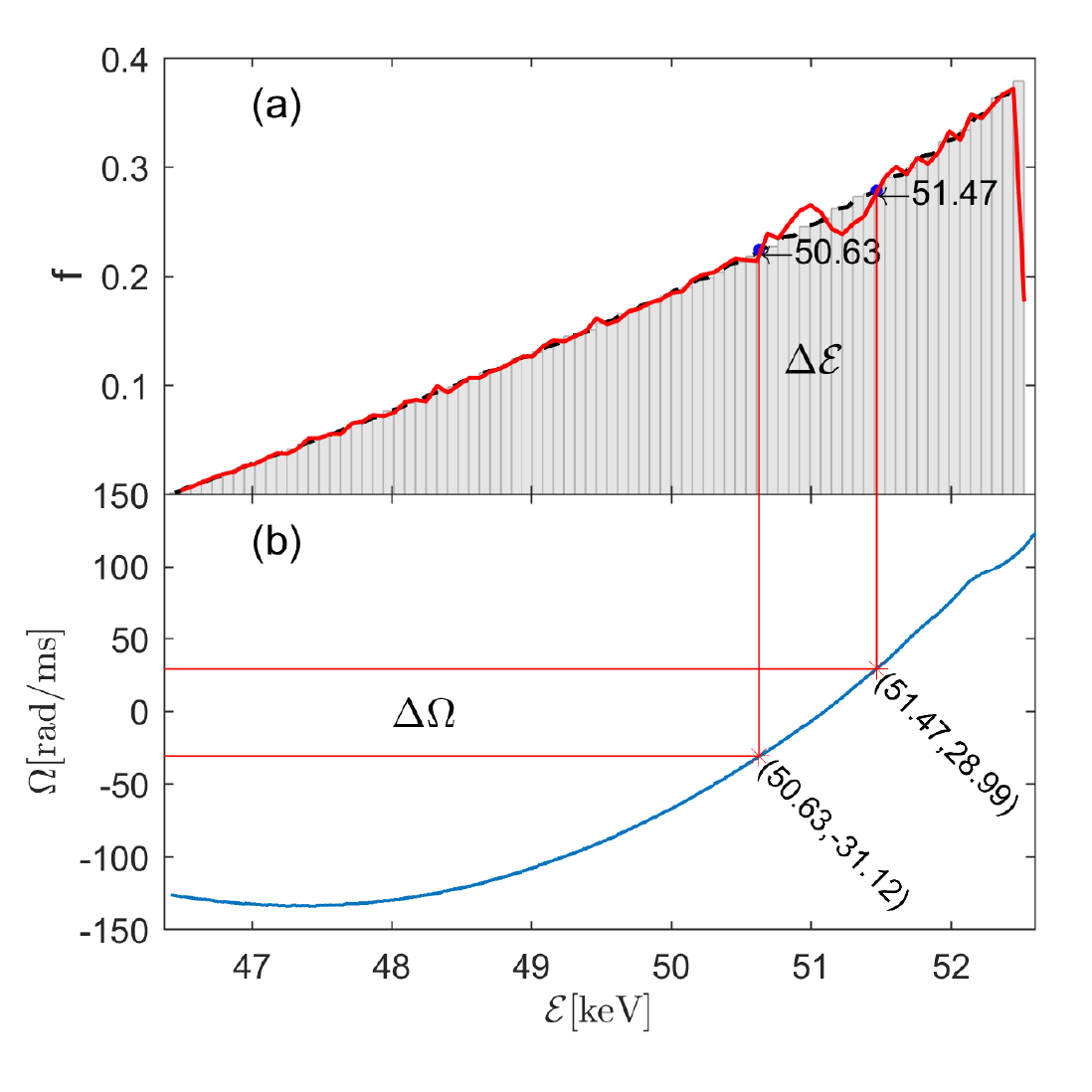}
	\caption{(a) The evolution of energy distribution: the dashed black line indicates the initial distribution with particle being loaded along the Res 2 line; the red line is the final distribution after a run time $T=3.89\;ms$ averaged over $\Delta t=0.078\;ms$, with the fixed mode amplitude $A=7.5\times10^{-5}$.
		(b) The function $\Omega(\mathcal{E})\big| _{\mathcal{E}'=45.09\;keV}$ shows the dependence of $\Omega$ (defined by Eq. \ref{eq:res}) as a function of energy along the Res 2 line. For this resonance, the resonance condition is $\Omega=4\langle\omega_\zeta\rangle-10\langle\omega_\theta\rangle-\omega_n$. 	
	}\label{fig:calRB}
\end{figure}

To calculate the resonance frequency broadening width $\Delta \Omega$, a new method is proposed to determine the broadening based on the linear distribution slope. $\Delta \Omega$ can be used to compute the broadening in $P_\zeta$. As mentioned above, the resonance broadening region is the platform that allows momentum and energy exchange between particles and waves. So we can measure the broadening of energy distribution $\Delta \mathcal{E}$ from the relaxation results, then calculate the resonance frequency broadening width $\Delta \Omega$ according to the function $\Omega(\mathcal{E})\big|_{\mathcal{E}'}$. An example of this procedure is shown in Fig. \ref{fig:calRB}. The EP relaxation in energy space is shown in Fig. \ref{fig:calRB}(a). The initially loaded 40 000 particles are divided into 80 bins, and the maximum number of particles in one bin is about 1000. The initial distribution of $\psi_p$ linearly decreases from $0.06$ to $0.37$, with $\theta_0=0$ and $\zeta_0$ being a random number. The initial distribution is almost linearly increasing in $P_\zeta$ according to Eq. \ref{eq:Pphi}. And the distribution in energy is the same with in $P_\zeta$ because $\mathcal{E}'$ is a constant. The total run time $T$ is about $3.89\;ms$ with a fixed mode, much larger than the trapping time, $T\gg\tau_b=2\pi/\omega_b$. Usually the particle distribution oscillates near the island, since energy of particles near separatrix oscillate with time. However, those oscillations do not contribute to the flattening. The relaxed distribution, shown in red line in the figure, is averaged over $\Delta t=0.078\;ms\ll\tau_b$ in order to remove these fluctuations. Figure \ref{fig:calRB}(b) shows the dependence of $\Omega$ (defined by Eq. \ref{eq:res}) as a function of energy along the Res 2 line, $\Omega(\mathcal{E})\big|_{\mathcal{E}'=45.09\;keV}$. For this resonance, the resonance condition is $\Omega=4\langle\omega_\zeta\rangle-10\langle\omega_\theta\rangle-\omega_n$.\\
\section*{5. Resonance Broadening due to a Single Mode}
\subsection*{5.1. Broadening Coefficient $a=\Delta\Omega/\omega_{b}$ in Realistic Cases}
\begin{figure*}\centering
	\subfigure
	{\label{fig:Res1df_Omg_E} 
		\includegraphics[width=3in, angle=0]{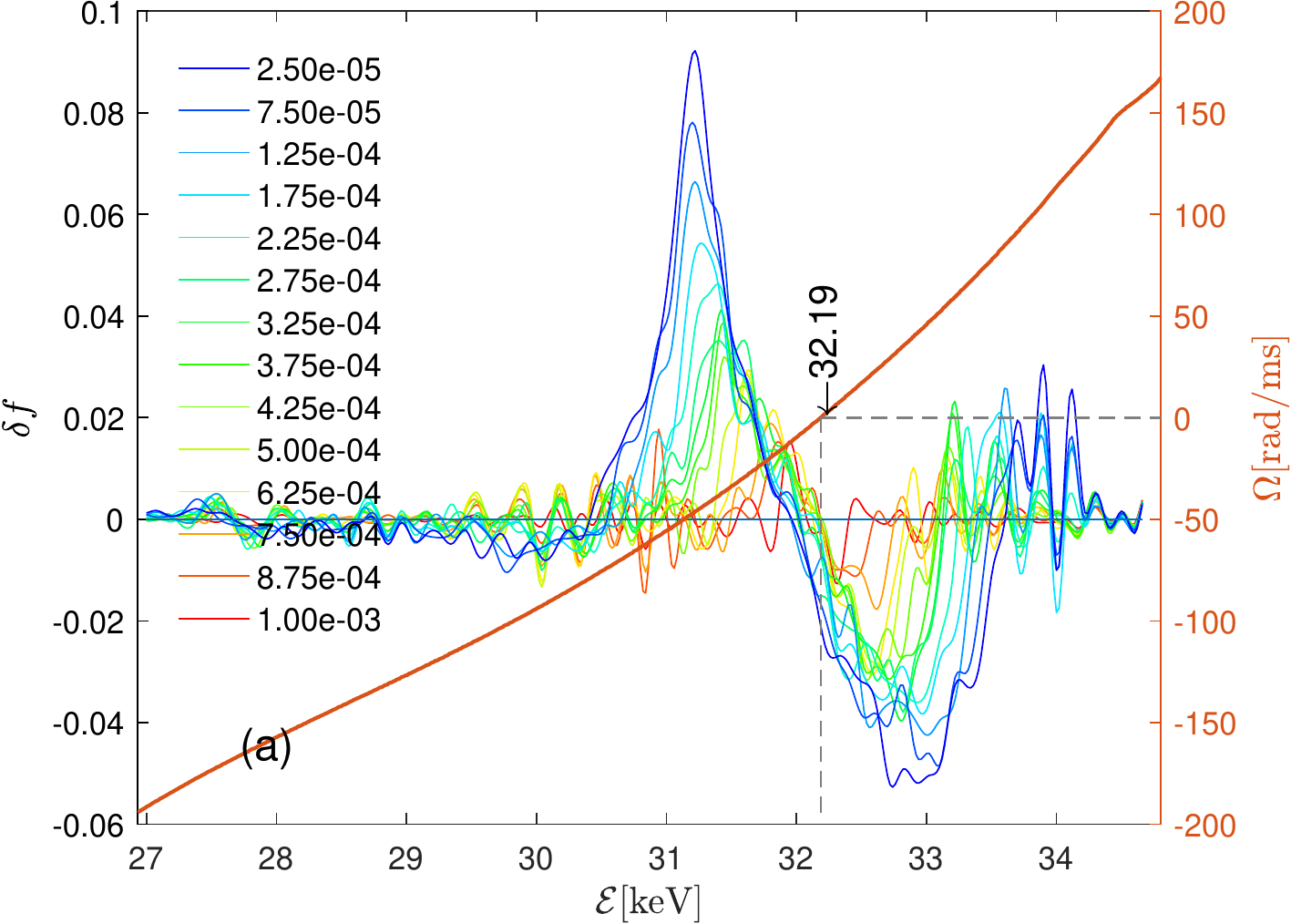}}
	\subfigure
	{\label{fig:Res1DOmg_wb} 
		\includegraphics[width=3in, angle=0]{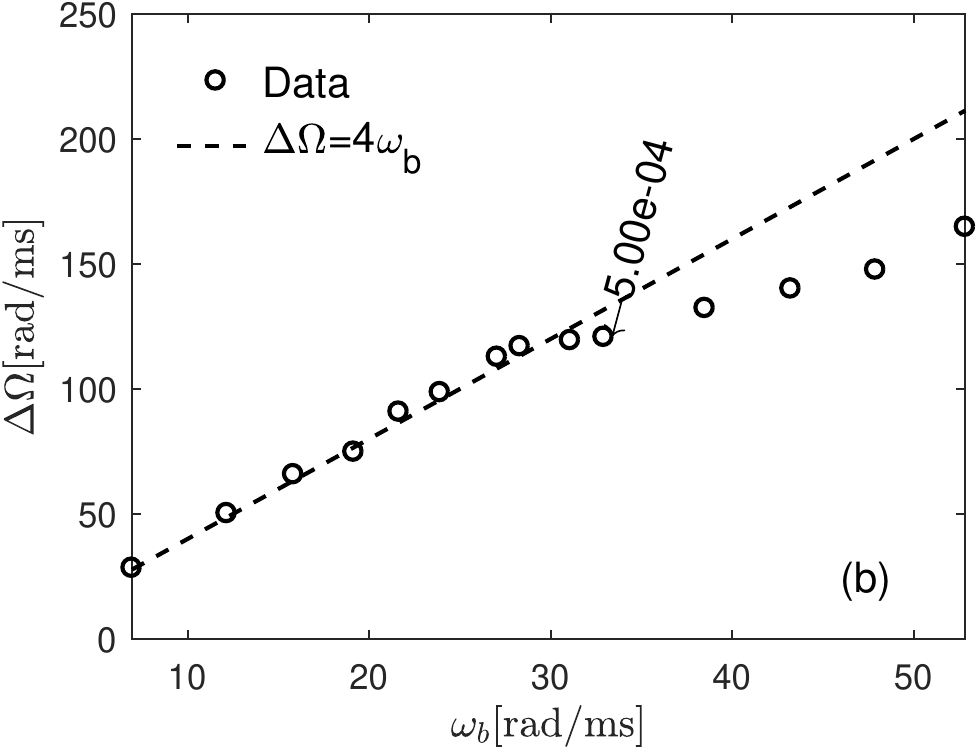}}
	\caption{Res 1 case. (a): Energy distribution change $\delta f$ and $\Omega(\mathcal{E})\big|_{\mathcal{E}'=30\;keV}$. (b): $\Delta\Omega$ variation with $\omega_b$. The value $5\times10^{-4}$ shown in the plot corresponds to the amplitude at the break down of the pendulum approximation.}
	\label{fig:Res1}                
\end{figure*}

\begin{figure*}\centering
	\subfigure 
	{\label{fig:Res2df_Omg_E} 
		\includegraphics[width=3in, angle=0]{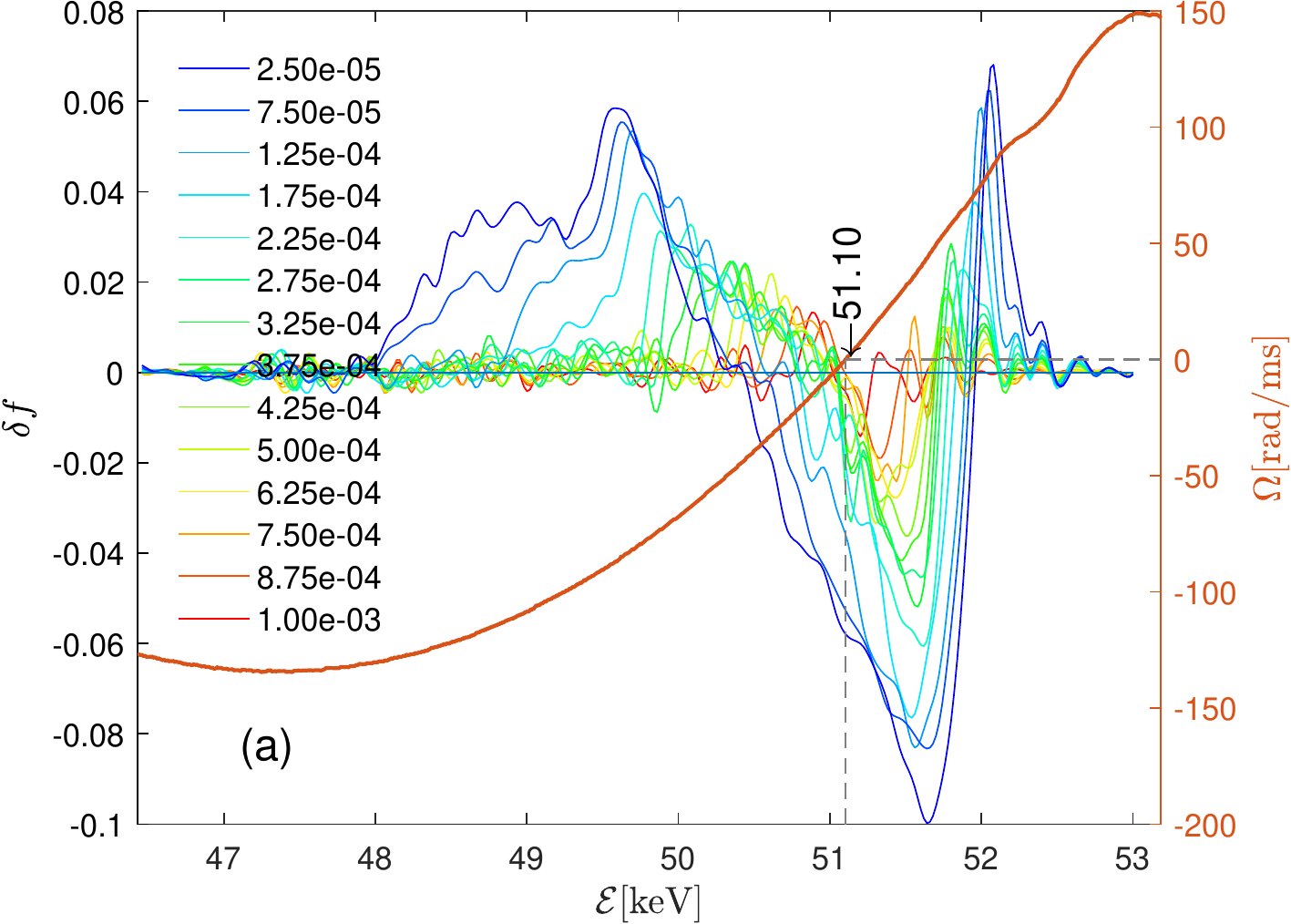}}
	\subfigure
	{\label{fig:Res2DOmg_wb} 
		\includegraphics[width=3in, angle=0]{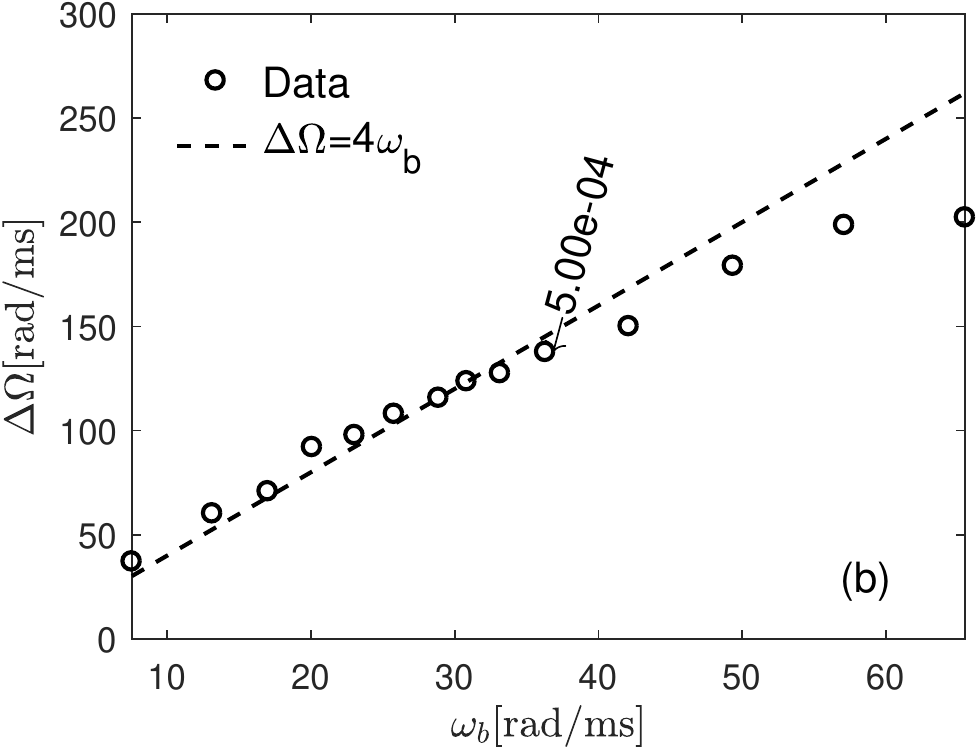}}
	\caption{Res 2 case. (a): Energy distribution change $\delta f$ and $\Omega(\mathcal{E})\big|_{\mathcal{E}'=45.09\;keV}$. (b): $\Delta\Omega$ variation with $\omega_b$.  The value $5\times10^{-4}$ shown in the plot corresponds to the amplitude at the break down of the pendulum approximation.}
	\label{fig:Res2}                
\end{figure*}

\begin{figure*}\centering
	\subfigure
	{\label{fig:Res3df_Omg_E} 
		\includegraphics[width=3in, angle=0]{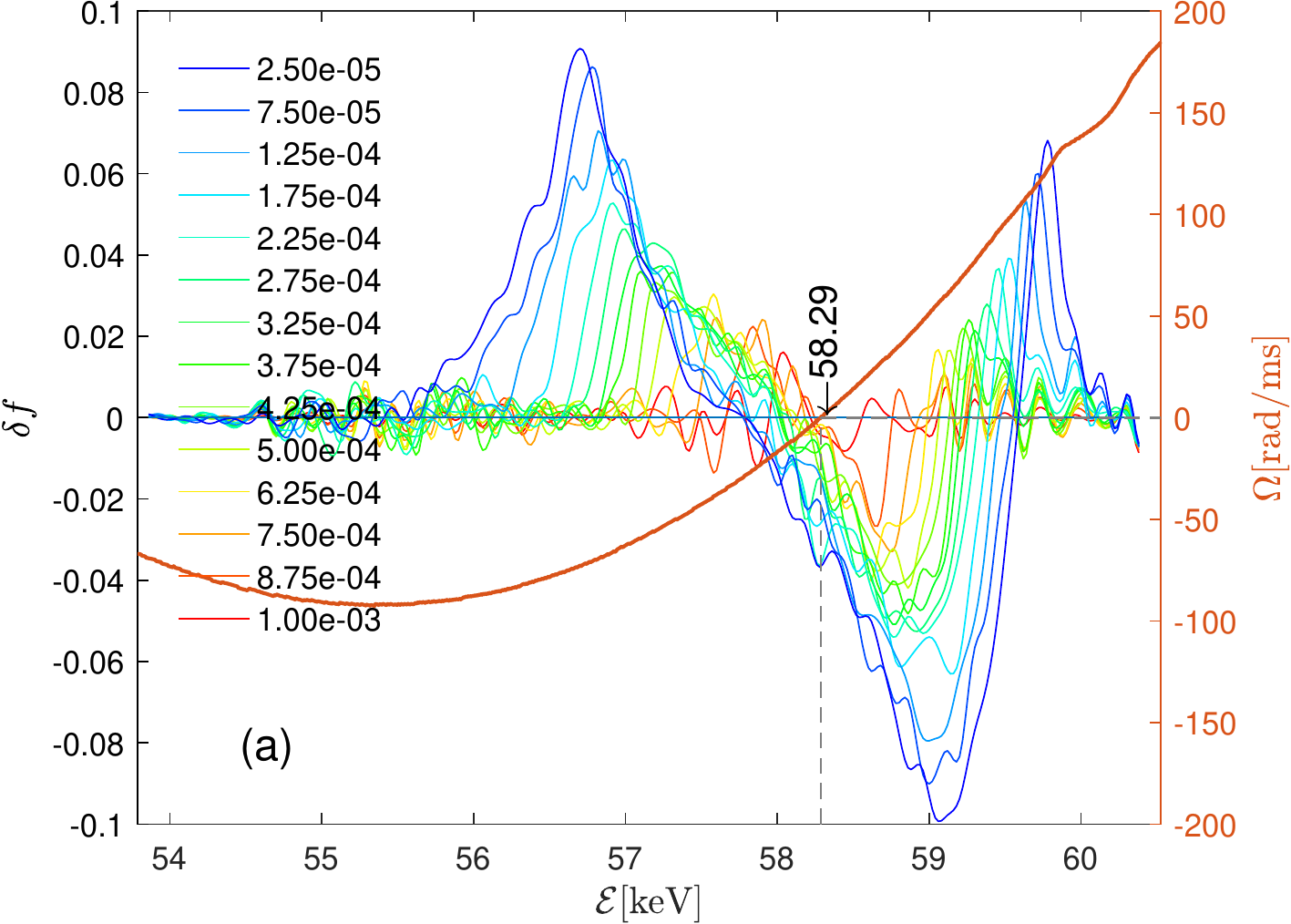}}
	\subfigure
	{\label{fig:Res3DOmg_wb} 
		\includegraphics[width=3in, angle=0]{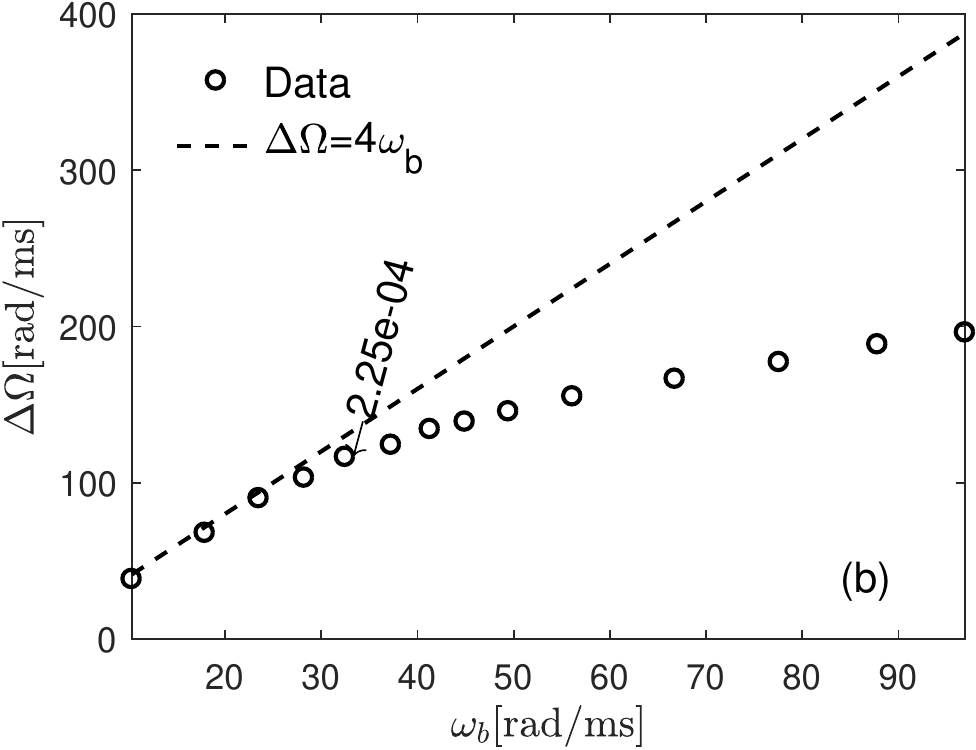}}
	\caption{Res 3 case. (a): Energy distribution change $\delta f$ and $\Omega(\mathcal{E})\big|_{\mathcal{E}'=51.93keV}$. (b): $\Delta\Omega$ variation with $\omega_b$. The value $2.25\times10^{-4}$ shown in the plot corresponds to the amplitude at the break down of the pendulum approximation.}
	\label{fig:Res3}                
\end{figure*}

\begin{figure*}\centering
	\subfigure
	{\label{fig:Res4df_Omg_E} 
		\includegraphics[width=3in, angle=0]{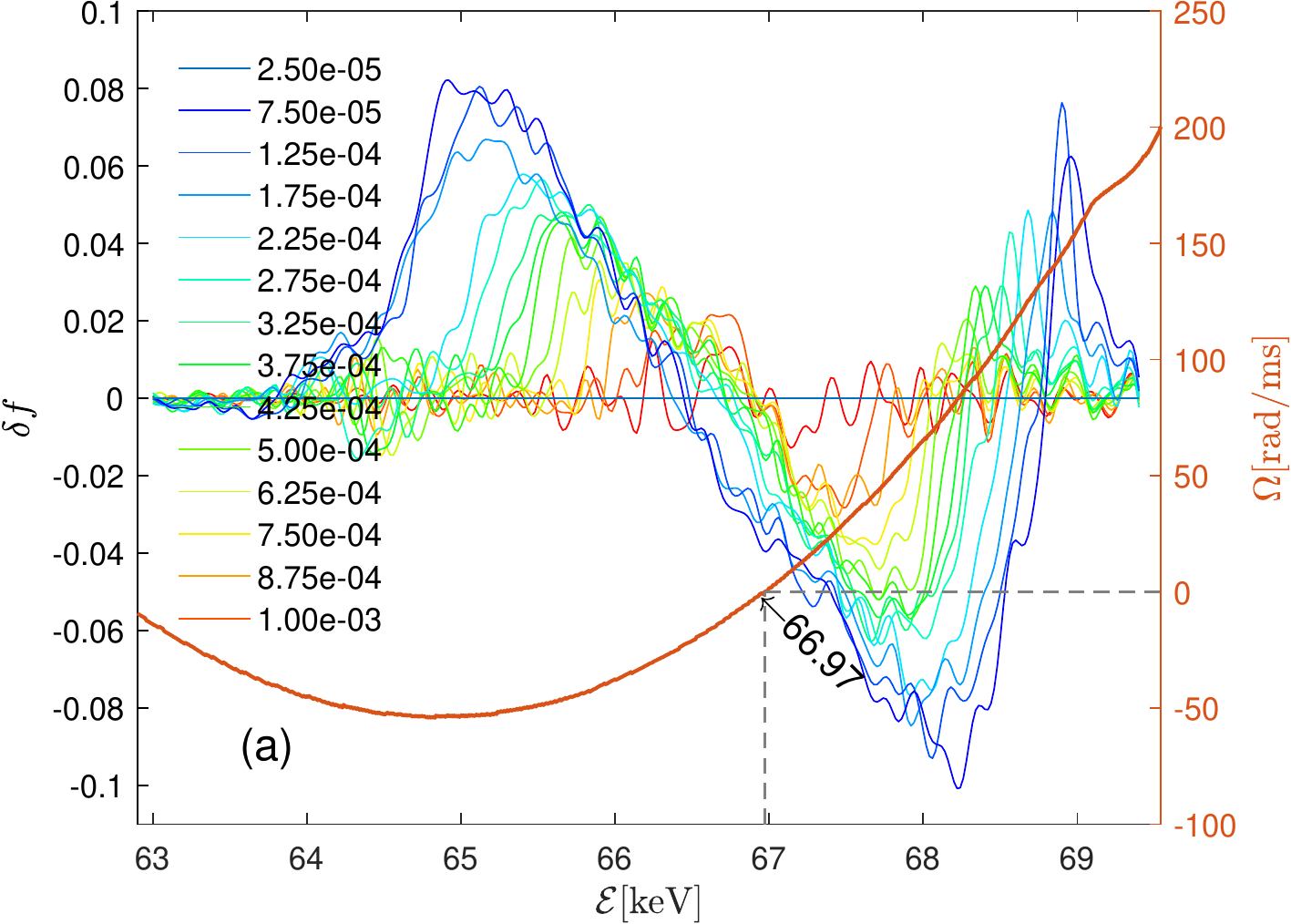}}
	\subfigure
	{\label{fig:Res4DOmg_wb} 
		\includegraphics[width=3in, angle=0]{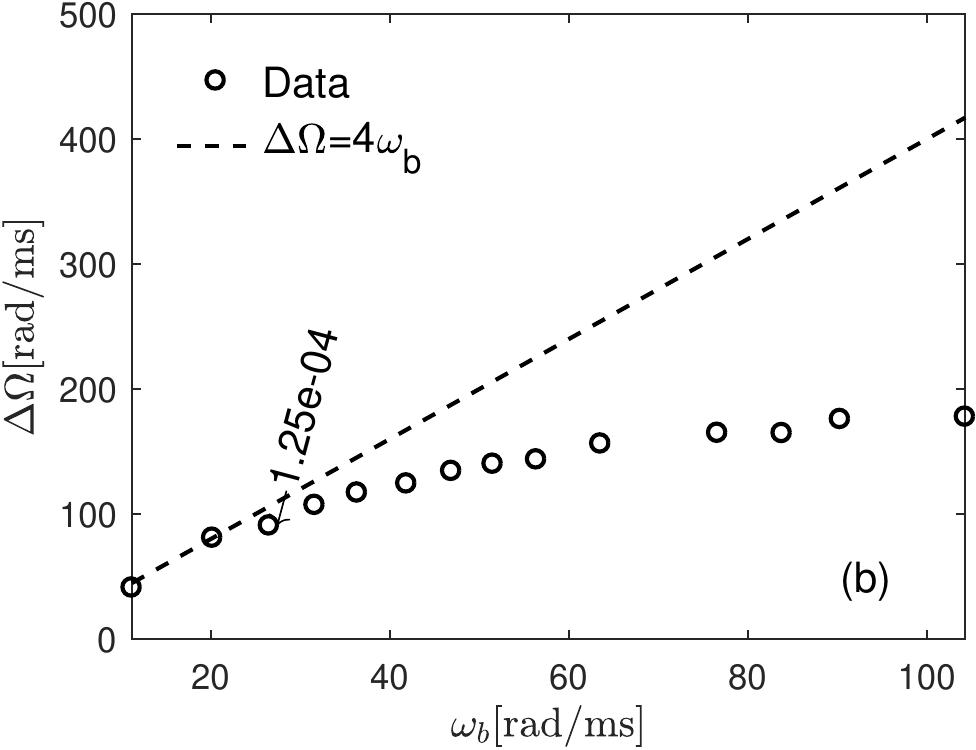}}
	\caption{Res 4 case. (a): Energy distribution change $\delta f$ and $\Omega(\mathcal{E})\big|_{\mathcal{E}'=60.22\;keV}$. (b): $\Delta\Omega$ variation with $\omega_b$. The value $1.25\times10^{-4}$ shown in the plot corresponds to the amplitude at the break down of the pendulum approximation.}
	\label{fig:Res4}                
\end{figure*}

\begin{figure}\centering
	\includegraphics[width=0.6\textwidth]{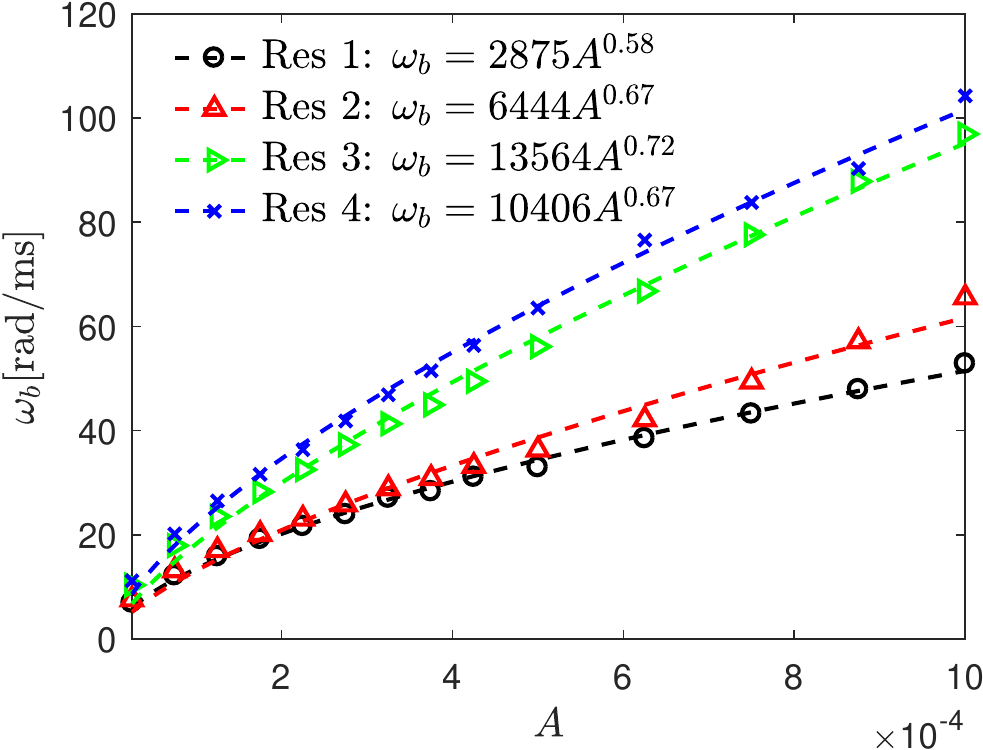}
	\caption{$\omega_{b}$ variation with mode amplitude $A$ at four cases. Dots are numerical data and dashed line are fitting results. At the same mode amplitude, the deeply trapped bounce frequency $\omega_{b}$ is increasing from Res 1 to 4.
	}\label{fig:allReswb_A}
\end{figure}

Following this process, we study the resonance broadening with different mode amplitudes for four resonance Res 1 to 4, as shown in Figs. \ref{fig:Res1} to \ref{fig:Res4}. \\
\indent The Res 1 case of Fig. \ref{fig:Res1} corresponds to particle initially loaded along the Res 1 line shown in Fig. \ref{fig:resPEplane}. Figure \ref{fig:Res1df_Omg_E} show the energy distribution change $\delta f=f(t=T)-f(t=0)$ with different mode amplitudes, with $\delta f>0$ on the left side of the island and $\delta f<0$ on the right. On the right hand side of Fig. \ref{fig:Res1} shows the function $\Omega(\mathcal{E})\big|_{\mathcal{E}'}$, the linear exactly resonant point $\mathcal{E}_{res}$ is also pointed out, $\mathcal{E}_{res}=32.19\;keV$. Figure \ref{fig:Res1DOmg_wb} shows the comparison of numerical results of $a=\Delta\Omega/\omega_b$ with the theoretical prediction of pendulum approximation, $\Delta\Omega/\omega_b=4$. When the mode amplitude $A<5\times10^{-4}$, numerical results agree with the pendulum approximation. Other cases are represented the same way.\\
\indent For Res 2, the linear resonant point is at $\mathcal{E}_{res}=51.1\;keV$ as shown in Fig. \ref{fig:Res2df_Omg_E}. When the mode amplitude $A<5\times10^{-4}$, numerical result agrees with the pendulum approximation as shown in Fig. \ref{fig:Res2DOmg_wb}.
For Res 3, the linear resonant point is at $\mathcal{E}_{res}=58.29\;keV$ as shown in Fig. \ref{fig:Res3df_Omg_E}. When the mode amplitude $A<2.25\times10^{-4}$, numerical results agree with the pendulum approximation as shown in Fig. \ref{fig:Res3DOmg_wb}.
For Res 4, the linear resonant point is at $\mathcal{E}_{res}=66.97\;keV$ as shown in Fig. \ref{fig:Res4df_Omg_E}. When the mode amplitude $A<1.25\times10^{-4}$, numerical results agree with the pendulum approximation as shown in Fig. \ref{fig:Res4DOmg_wb}.\\
\indent There are several similarities in these cases. The deeply trapped bounce frequency is almost proportional to the square root of the mode amplitude, $\omega_b\propto\sqrt{A}$, as shown in Fig. \ref{fig:allReswb_A}. It is found that the coefficient $a$ decreases as mode amplitude increases for these four cases, which indicates that the pendulum approximation breaks down. For Res 3 and Res 4, the pendulum approximation breaks down at lower mode amplitude in comparison with Res 1 and 2. This is because that at the same mode amplitude, the island width and the deeply trapped bounce frequency are larger for these four cases as shown in Fig. \ref{fig:poinkall} and \ref{fig:allReswb_A}. 
It is also found that the pendulum approximation breaks down at $\omega_{b}$ around $30\;rad/ms$ from Figs. \ref{fig:Res1DOmg_wb} to \ref{fig:Res4DOmg_wb}.\\
\indent It is found that the platform center does not move significantly for Res 1 case. The derivative $\partial \Omega/\partial\mathcal{E}$ is roughly the same at $\mathcal{E}<\mathcal{E}_{res}$ and $\mathcal{E}>\mathcal{E}_{res}$. For Res 2, 3 and 4 cases, the platform center moves to the left significantly and the energy broadening $\Delta\mathcal{E}$ does not increase equally on each side. The derivative $\partial \Omega/\partial\mathcal{E}$ is smaller at the $\mathcal{E}<\mathcal{E}_{res}$ region. So the resonance islands of Res 2, 3 and 4 are highly asymmetric compared to Res 1 as shown in Fig. \ref{fig:poinkall}, and the broadening in energy is larger at $\mathcal{E}<\mathcal{E}_{res}$.

\subsection*{5.2. Pendulum Approximation Viability for Realistic cases}
\begin{figure*}\centering
	\subfigure
	{\label{fig:Res1wballamp} 
		\includegraphics[width=3in, angle=0]{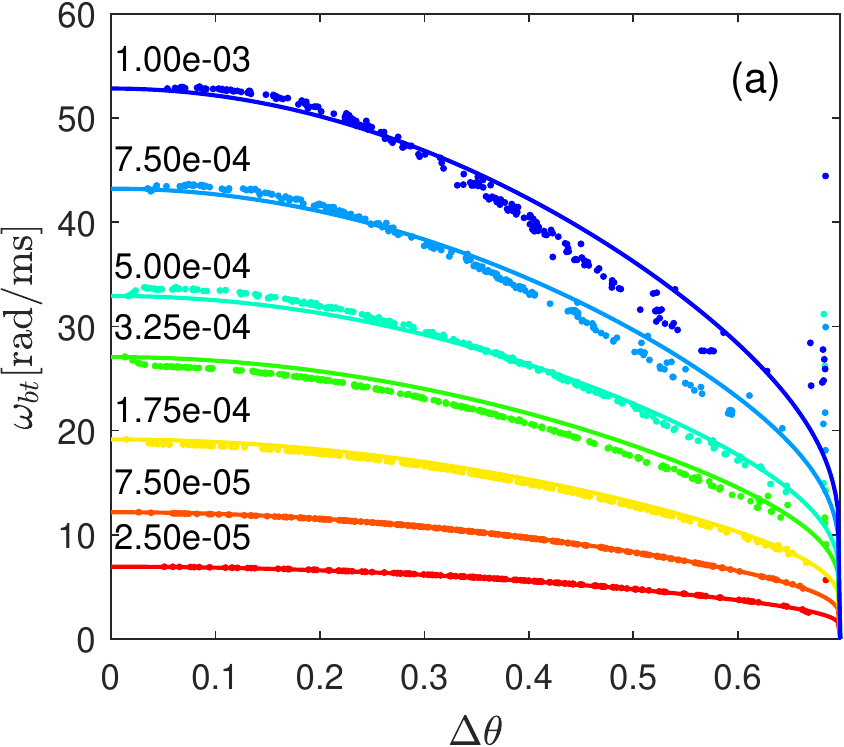}}
	\subfigure
	{\label{fig:Res1ratioA}
		\includegraphics[width=3in, angle=0]{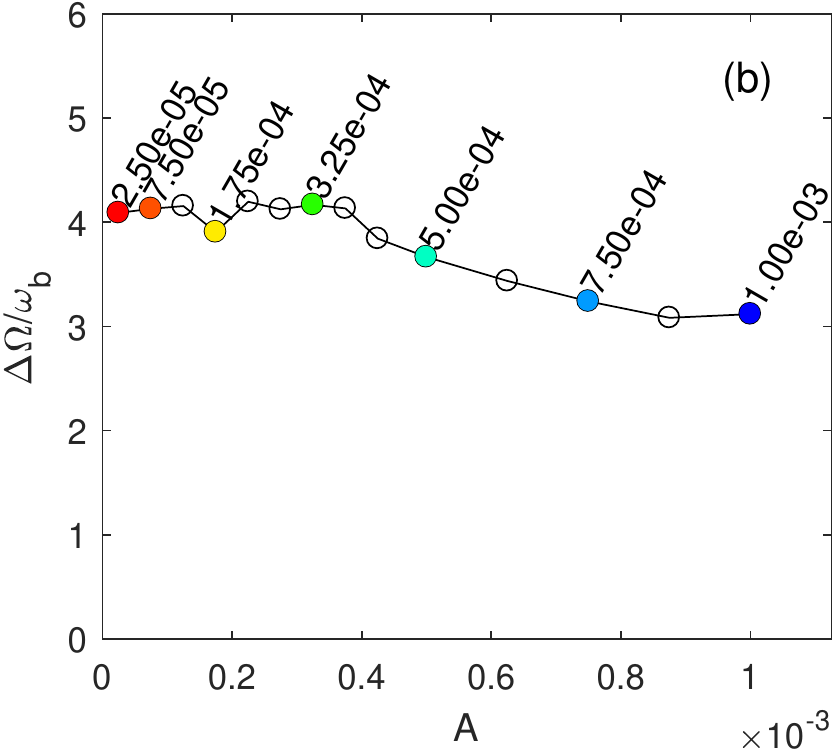}}
	\subfigure
	{\label{fig:Res3wballamp} 
		\includegraphics[width=3in, angle=0]{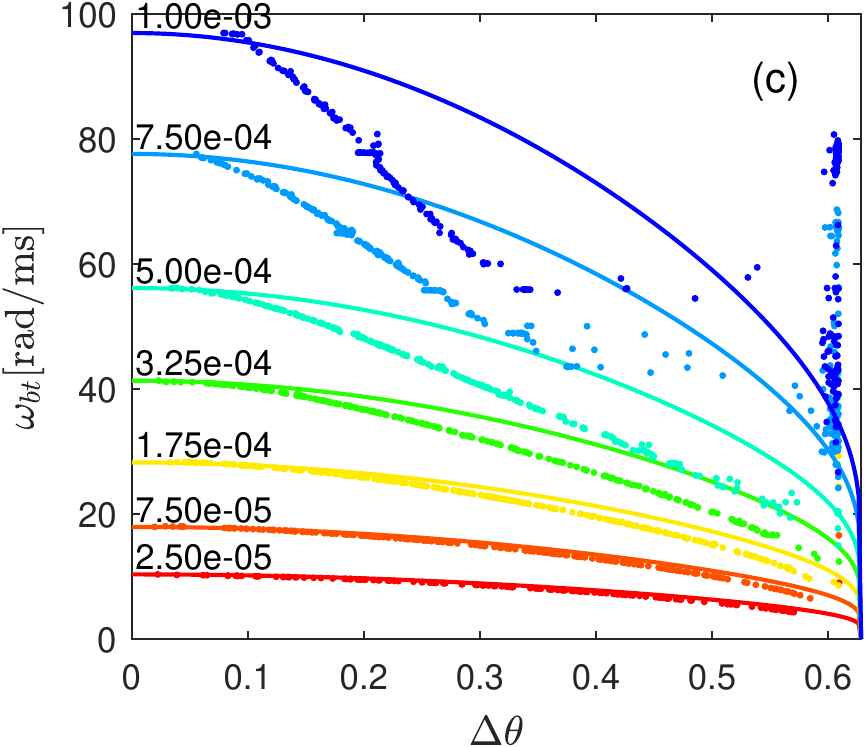}}
	\subfigure
	{\label{fig:Res3ratioA} 
		\includegraphics[width=3in, angle=0]{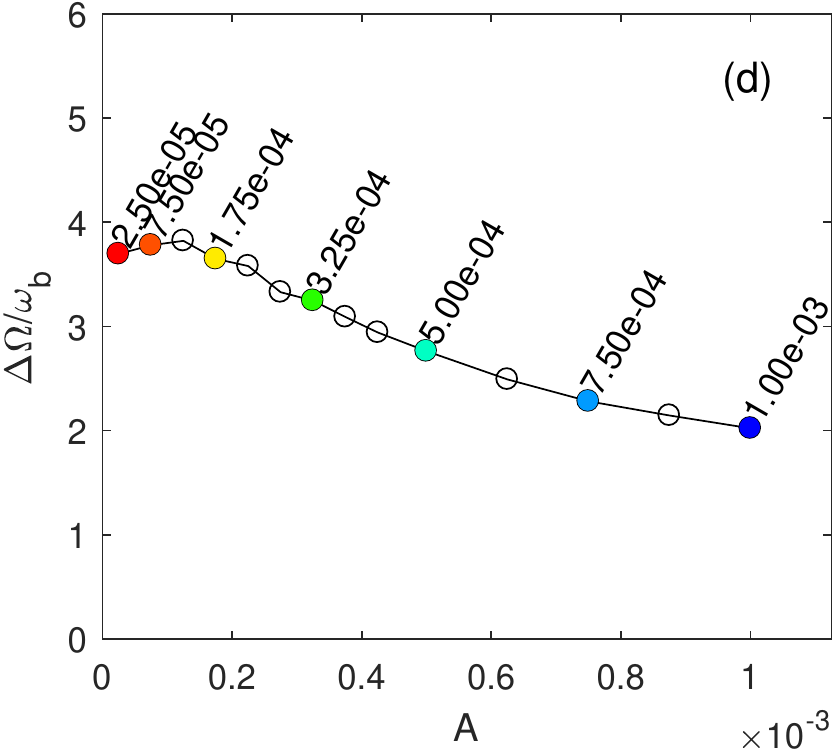}}
	\caption{(a) and (c): Trapping frequency variation with $\Delta\theta$. Solid lines are fitting curves from the analytical prediction shown in Eq. \ref{eq:wbt}, points are trapping frequencies calculated numerically. (b) and (d): Coefficient $a=\Delta\Omega/\omega_b$ variation with different mode amplitudes $A$. Figures (a) and (b) are for Res 1 ($l=9$): particle trapping frequency $\omega_{bt}$ agrees with theory, and $a\approx4$ for all mode amplitudes. Figures (c) and (d) are for Res 3, ($l=10$): $a$ decreases with mode amplitude A. When $A\geq3.25\times10^{-4}$, the pendulum approximation breaks down substantially.}
	\label{fig:pendcoeffa}                
\end{figure*}

\begin{figure*}\centering
	\subfigure
	{\label{fig:amp1} 
		\includegraphics[width=1.9in, angle=0]{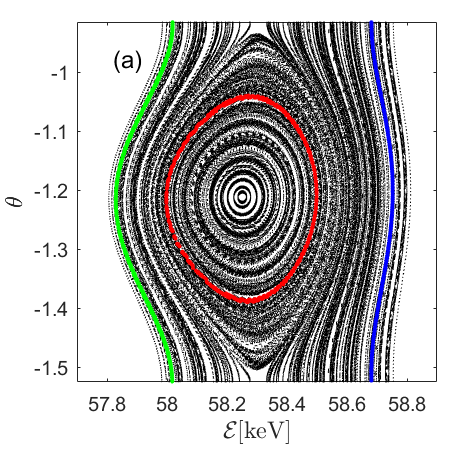}}
	\subfigure
	{\label{fig:deform} 
		\includegraphics[width=1.9in, angle=0]{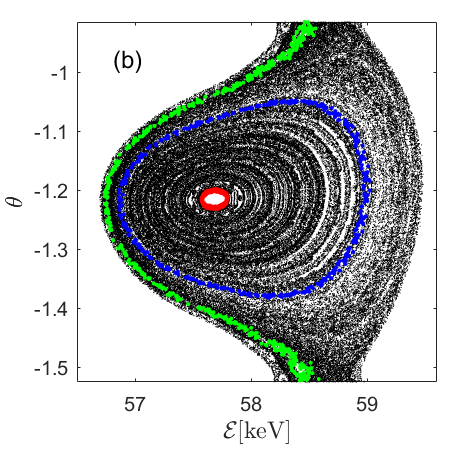}}
	\subfigure
	{\label{fig:chaos}
		\includegraphics[width=1.9in, angle=0]{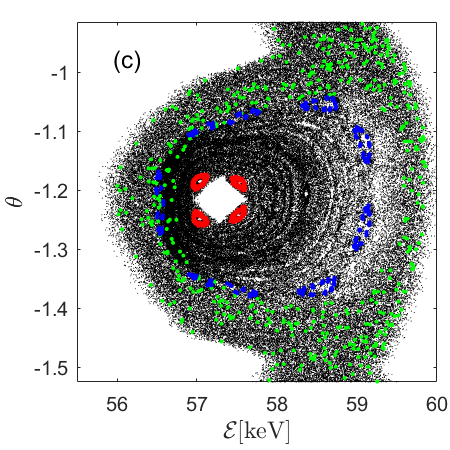}}
	\caption{The process of pendulum approximation breaking down for Res 3. The orbit of one particle with same initial coordinates $(\psi_{p0},\theta_0,\zeta_0,\rho_{\parallel 0})$ is highlighted in the same color. (a): At $A=2.5\times10^{-5}$, the trapping frequency agrees with the theory. (b): At $A=5\times10^{-4}$, the resonance island shows deformation. (c): At $A=1\times10^{-3}$, the separatrix becomes a band of chaotic trajectories.}
	\label{fig:breakdown}                
\end{figure*}

As mentioned above, the pendulum approximation works very well up to a certain point of the mode amplitude in realistic cases. It is found that $a$ is getting increasingly smaller than 4 as the mode amplitude increases, and the pendulum prediction of the coefficient $a$ breaks down at smaller mode amplitude for Res 3 compared to Res 1. So we investigate the pendulum approximation at Res 1 and Res 3, as shown in Fig. \ref{fig:pendcoeffa}.\\ 
\indent The solid lines in Figs \ref{fig:Res1wballamp} and \ref{fig:Res3wballamp} are fitting curves using Eq. \ref{eq:wbt}, with the $\omega_b$ set as the numerical trapping frequency of particle with the minimal $\Delta\theta$, and points are trapping frequencies calculated numerically. Some points are vertically aligned at the same $\Delta\theta$ near the right edge because of the particle stochastic motion at the separatrix. For Res 1, the particle trapping frequency $\omega_{bt}$ agrees with the theory as shown in Fig. \ref{fig:Res1wballamp}, and $a=\Delta\Omega/\omega_b\approx4$ as shown in Fig. \ref{fig:Res1ratioA} up to mode amplitude $A=5\times10^{-4}$. For Res 3, when the mode amplitude $A=3.25\times10^{-4}$, it can be clearly seen that the trapping frequency deviates from the pendulum approximation as shown in Fig. \ref{fig:Res3wballamp}. When the mode amplitude $A=7.5\times10^{-4}$, the numerical trapping frequencies do not show a regular pattern when $\Delta\theta>0.3$ because particle orbits become chaotic in that regime.\\
\indent We observe that the trapping frequency is always below the theory predicted by Eq. \ref{eq:wb}, e.g., when the mode amplitude $A=3.25\times10^{-4}$ at Res 3. It is found that the resonance island is highly asymmetric in this case, as shown in Fig. \ref{fig:deform}. When the mode amplitude is large, e.g., $A=10^{-3}$, the separatrix no longer exists and becomes a band of chaotic trajectories with nonzero thickness as shown in Fig. \ref{fig:chaos}.\\
Equation \ref{eq:pendulum} describes an exact pendulum and gives rise to an exactly defined separatrix between the island interior and the topologically unperturbed exterior. However, for the realistic cases, the amplitude of the wave trapping well is related with the poloidal flux, i.e., the wave is non-sinusoidal, the WPI term in Eq. \ref{eq:pendulum} is no longer sinusoidal function with a constant amplitude $\omega_b^2$. In addition, there is not only one resonance and other resonances will give an additional perturbation term to Eq. \ref{eq:pendulum}. Then the trapped particle motion can become highly nonlinear and more complicated \cite{white2013TCP}
\begin{equation}\label{eq:exactmotion}
\frac{d^2\alpha}{dt^2}+V(\psi_p)sin(\alpha-\omega_{n}t-\alpha_0)+\epsilon sin(j\alpha-\omega_{n}t-\varphi_0)=0.
\end{equation}
When the mode amplitude becomes large, the resonance island becomes too wide and the eigenmode structure is varying within the resonance island so using a constant $V$ is not a good approximation anymore. The third term of Eq. \ref{eq:exactmotion} accounts for small high-order oscillations, with $j$ an integer. In principle, we can find the deviation of the trapping frequency and estimate the thickness of the chaos band. However, there are no analytic representations for $V(\mathcal{E})$ and $\epsilon sin(j\alpha-\omega_{n}t-\varphi_0)$.\\
\indent We choose three representative mode amplitudes of the Res 3 case and plot the resonance islands to discuss the process and reasons of pendulum approximation break-down: at $A=2.5\times10^{-5}$, numerical trapping frequency $\omega_{bt}(\Delta\theta)$ agrees with pendulum prediction in Eq. \ref{eq:wbt}; at $A=5\times10^{-4}$, $\omega_{bt}(\Delta\theta)$ drops below the prediction; at $A=1\times10^{-3}$, $\omega_{bt}(\Delta\theta)$ is irregular, as shown in Fig. \ref{fig:Res3wballamp}. We observe that the trapping frequency shifts down with respect to the theory first due to the resonance island deformation, then the particle orbit becomes stochastic at large mode amplitude as shown in Fig. \ref{fig:breakdown}. When the mode amplitude is small, $A=2.5\times10^{-5}$, the resonance island shape looks like the exact pendulum as shown in Fig. \ref{fig:amp1}. As the mode amplitude increases, $A=5\times10^{-4}$, the resonance island deforms and stretches along the $\mathcal{E}$ direction as shown in Fig. \ref{fig:deform}. When the mode amplitude $A=1\times10^{-3}$, significant chaos occurs as shown in Fig. \ref{fig:chaos}. Particle orbits near the ``separatrix'' become a thick chaos band, as an example, one particle orbit is highlighted in green. Actually, there is no defined ``separatrix'' because there are always a series of higher-order perturbed terms in the Hamiltonian due to the complexity of the realistic equilibrium. But at small mode amplitude, the perturbed Hamiltonian can be neglected and the chaos band is very narrow. In addition, an example of particle orbit change due to higher order resonance is highlighted in red. The linear $l=10$ island is surrounded by 4 small islands, which means $l=40$ of this higher order resonance. Another particle orbit is highlighted in blue with $l=70$. These high order perturbations have an important effect on trapped particle motion when the mode amplitude is large.   
\subsection*{5.3. Averaging in Constant of Motion Phase-space}
When the mode amplitude is small, the coefficient $a=4$ is a good choice to compute the resonance line broadening. When the mode amplitude becomes larger, the coefficient is observed in the simulation to be smaller than 4 and dependent on the resonance location in the COM phase space. As inferred from Fig. \ref{fig:Res1DOmg_wb} to \ref{fig:Res4DOmg_wb}, Fig. \ref{fig:wormamp6} and Fig. \ref{fig:pendcoeffa}, the pendulum approximation breaks down at the dominant resonance (at the position of largest perturbation) and higher energy regime of phase-space when the mode amplitude $A=5\times10^{-4}$ but still works well at a large domain where the resonance islands are narrow, which means small $\omega_{b}$. Figure \ref{fig:poinkall} and \ref{fig:allReswb_A} show that, for the case of a fixed mode amplitude, the resonant islands become broader for larger resonant energies, meaning that also the bounce frequency should increase, which is consistent with expression Eq. \ref{eq:wb}. 
Complementary to the energy dependence of the bounce frequency of the deeply-trapped particles, we see from Fig. \ref{fig:pendcoeffa} that, for a given mode amplitude, the pendulum approximation is observed to be poorer for higher resonant energies. In addition to the nonlinear dynamics dependence on the energy, we also examined the effect of the eigenstructure. We show in Fig. \ref{fig:resPEplane} the curve (in dashed black line) representing the peak of the mode structure, which represents particle orbit passing through $(\psi_p=0.22,\theta=0)$ point. We note that the bounce frequency Eq. \ref{eq:wb}, through its dependence on the electric field, is also sensitive to the radial spread of the mode, being larger for cases in which the resonance location is closer to the mode structure peaks location. Since we are aiming to find the universal coefficients, the averaging of broadening coefficients in the COM phase-space is recommended \cite{nikolai99saturation,duarte2017onset}.\\
\begin{equation}
\overline{a}=\int a Q  d\Gamma \bigg/ \int Q d\Gamma
\end{equation}
where $d\Gamma$ is the differential phase-space volume, $Q \sim \xi^2\delta(\Omega) \sim \omega_b^4\delta(\Omega)$, the $\delta$ function gives the integration along the resonance curve.

\section*{6. Summary}
In this work, we study the resonance broadening due to a single wave in a realistic DIII-D plasma. We propose a numerical approach to measure the width of the resonance for the distribution function flattening and to calculate the corresponding resonance broadening $\Delta\Omega$. The evolution of the distribution function is obtained by using the guiding center, test particle code ORBIT for DIII-D discharge 159243. In principle, this method can be employed in different codes to measure the resonance broadening for cases in which the resonant particle relaxation is known. We also propose a numerical technique and develop a MATLAB package to find the trapped particles and calculate the trapping frequency corresponding to WPI. This method can in general be used in different codes for analyses of both configuration-space (such as \Poincare plots of magnetic islands) and phase-space island properties.\\  
\indent Our work studies the validity of WPI in tokamaks being formally described as a nonlinear pendulum. It is found that when the mode amplitude is small, the nonlinear pendulum approximation is applicable and the coefficient $\Delta\Omega/\omega_b$ equals to 4 as analytically predicted (in Sec. 4). As the mode amplitude increases, the resonance island deforms, and then significant chaos takes place, bringing the realistic case gradually away from the pendulum approximation validity. It is found that the trapping frequency drops down faster than the theory predicts. Consequently, the calculated coefficient becomes smaller than 4. 
\section*{Acknowledgments}
The authors would like to thank Prof. Guoyong Fu for helpful discussions. We also thank Dr. Cami Collins and Dr. Mario Podesta for providing the DIII-D parameters used in this work. This work was mainly supported by the Science Challenge Project Grant No. JCKY2016212A505, NSFC (Grant Nos. 41674165 and 11261140326) and ITER-CN programs (No. 2014GB107004 and 2013GB11001). This work was also performed under the auspices of the U. S. Department of Energy (DOE) under contracts DE-AC02-09CH11466 and DE-FC02-04ER54698. GM thanks the China Scholarship Council (CSC), as the work was done during her visit to PPPL sponsored by CSC. VND was partially supported by the S\~ao Paulo Research Foundation (FAPESP, Brazil) under grants 2012/22830-2 and 2014/03289-4.
\section*{Appendix A}
We propose a numerical method to calculate the trapping frequency and developed a MATLAB package for ORBIT kinetic \Poincare plot data analysis. This method is general for particle orbit classification and trapping frequency calculation and can be used for data analysis for other codes. 
\subsection*{A.1. Search for Resonance Island}
To get a high-resolution kinetic \Poincare of the resonance island for calculating the trapping frequency changing with $\Delta\theta$, the first step is to give initial particle guiding center coordinates $(\psi_p,\theta,\zeta,\rho_{\parallel})$ and to load the particles uniformly inside the resonance island. For the sake of simplicity, the following approach is adopted. The first step is to use the resonance frequency plot Fig. \ref{fig:OmegaPEplane} or the phase vector rotation plot Fig. \ref{fig:wormamp6} to first roughly find the resonance location $(\mathcal{E},P_\zeta)$. In principle, the $(\psi_p,\theta,\rho_{\parallel})$ can be calculated accordingly. However there are no analytic equations to obtain them easily since $B(\psi_p,\theta)$ is not prescribed analytically, so we load particles uniformly in $(\psi_{p1},\psi_{p2})$ instead of loading particles uniformly on the line of $(\mathcal{E},P_\zeta)$ with constant $\mathcal{E}'$ with initial poloidal angle $\theta=0$. Choosing $\theta$ does not restrict the mode-particle phase since the toroidal angle $\zeta$ is a random number from 0 to $2\pi$ \cite{white2013TCP}. Now we just need to find the $(\psi_{p1},\psi_{p2})$ of the resonance. In equilibrium, particle orbits which pass through the $(\psi_{p},\theta=0)$ point satisfy
\begin{equation}
\frac{((P_\zeta+\psi_p)B(\psi_p,0))^2}{2g(\psi_p)^2}+\mu B(\psi_p,0)-\mathcal{E}=0.
\end{equation}
It is a parabola on the $(\mathcal{E},P_\zeta)$ plane, similar to the magnetic axis line as shown in Fig. \ref{fig:resPEplane}. Therefore we can find the $(\psi_{p1},\psi_{p2})$ corresponding to the resonance. We load 300 particles uniformly in a range of $0.04\psi_w$ which is enough to show details of the resonance island structure.
\subsection*{A.2. Particle Orbit Classification}
Usually the resonance island is highly asymmetric in $\mathcal{E}$ direction but nearly symmetric in $\theta$ direction, so we choose $\Delta\theta$ as the variable to represent the distance from the O-point. We select one resonance island in the island chains on the \Poincare plots by selecting the points only in a region of $\theta$. The boundary is set at two X-points $[\theta_L,\theta_U]$ by hand. The selected $\theta$ range is slightly smaller than the island length, $\theta_U-\theta_L\approx2\pi/l$. Then we compute the orbit length $\Delta\theta$ and width $W_{bt}=\mathcal{E}_{max}-\mathcal{E}_{min}$ of every particle. Note that, there are always some passing particle orbits near the resonance island as shown in Fig. \ref{fig:island}, their $\Delta\theta$ always equals to the $\theta_U-\theta_L$. Using $\Delta\theta$ can not tell how far from O-point a passing particle is. The pseudo trapping frequency of these passing particles will give some points vertically aligned at $\Delta\theta=\theta_U-\theta_L$. So we find a way to classify the particle orbit type to be either trapped or passing.\\
\indent A particle with the smallest $\Delta\theta$ is marked as the most deeply trapped particle, which gives $\omega_b$. A particle with $\Delta\theta\leq0.95(\theta_U-\theta_L)$ is identified as trapped, then we need to classify the passing and trapped particles near separatrix. The particle which has the maximum energy change, i.e., the largest orbit width $W_{bt}$, is at separatrix. The width of its trajectory in $(\mathcal{E}, \theta)$ space is identified as the separatrix width $W_{Sp}$. Then particles with $W_{bt}>0.95W_{Sp}$ are identified as trapped ones to find the trapped particle with $\Delta\theta>0.95(\theta_U-\theta_L)$. All the others are identified as passing.
\subsection*{A.3. Trapping Frequency Calculation}
The action angle $\theta$ change with time for a trapped particle is basically sinusoidal plus a noise level. We use Fast Fourier Transform (FFT) to calculate the trapping frequency automatically. It can find the dominant frequency of the signal $\theta(t)$, which is the trapping frequency of the particle, which thus improves the robustness of this method when the action angle boundaries $[\theta_L,\theta_U]$ are not selected accurately and when the particle orbit is influenced by higher order resonances. The run time is chosen to be about $T=10\;ms$. The number of sampling points of the signal data is about 500, which are interpolated to 2048. The FFT calculates discrete frequency with accuracy proportional to $1/T$. In order to smooth out aliasing errors, the following Gaussian filter is also adapted 
\begin{equation}
\mathcal{F}_{G}=\frac{1}{\sqrt{\pi}T_G erf(\frac{T}{2T_G})}exp\left[ -\left( \frac{t-T/2}{T_G}\right) ^2\right],
\end{equation}
where $erf(x)=\frac{2}{\sqrt{\pi}}\int_0^x e^{-t^2} dt$ is the error function. The $T_G$ of the Gaussian filter is chosen as $T_G=T/4$, which is much larger than the bounce period $\tau_b$. The trapping frequency is calculated from the weighted mean value of frequencies corresponding to the proper amount of the dominant Fourier components. When particle moves stochastically such as at the separatrix, the numerical trapping frequency varies because the signal $\theta(t)$ will include points actually from up and down resonance islands and increase the numerical trapping frequency as shown in Fig. \ref{fig:Res1wballamp} and \ref{fig:Res3wballamp}.
\providecommand{\newblock}{}

\end{document}